\documentclass[prd,aps,twocolumn,superscriptaddress,nofootinbib,preprintnumbers,floatfix]{revtex4-1}

\usepackage[bookmarks=false]{hyperref}
\usepackage{amsmath}
\usepackage{graphicx}
\usepackage[small]{subfigure}
\usepackage{placeins}
\usepackage{xcolor}
\usepackage{slashed}

\newcommand{\eq}[1]{Eq.~\eqref{eq:#1}}

\renewcommand{\sec}[1]{Sec.~\ref{sec:#1}}

\newcommand{\subsec}[1]{Sec.~\ref{subsec:#1}}

\newcommand{\fig}[1]{Fig.~\ref{fig:#1}}
\newcommand{\figs}[2]{Figs.~\ref{fig:#1} and \ref{fig:#2}}

\newcommand{\ord}[1]{\mathcal{O}(#1)}

\newcommand{\df}{\mathrm{d}}

\newcommand{\eps}{\epsilon}

\newcommand{\cO}{{\mathcal O}}

\newcommand{\bn}{\bar{n}}

\newcommand{\GeV}{\,\mathrm{GeV}}
\newcommand{\TeV}{\,\mathrm{TeV}}

\newcommand{\nn}{\nonumber}

\newcommand{\Ecm}{E_\mathrm{cm}}

\newcommand{\nons}{\mathrm{nons}}

\newcommand{\sub}{\mathrm{sub}}
\newcommand{\LO}{\mathrm{LO}}
\newcommand{\hadcm}{\mathrm{had\,cm}}

\newcommand{\Tau}{\mathcal{T}}

\newcommand{\cut}{{\mathrm{cut}}}

\allowdisplaybreaks[4]

\begin{document}

%%%%%%%%%%%%%%%%%%%%%%%%%%%%%%%%%%%%%%%%%%%%%%%%%%%%%%%%%%%%%%%%%%%%%%%%%%%%%%%%
% Title page
%%%%%%%%%%%%%%%%%%%%%%%%%%%%%%%%%%%%%%%%%%%%%%%%%%%%%%%%%%%%%%%%%%%%%%%%%%%%%%%%

\preprint{\vbox{\hbox{MIT--CTP 4939}\hbox{DESY 17-128}}}

\title{\boldmath N-Jettiness Subtractions for $gg\to H$ at Subleading Power}

\author{Ian Moult}
\affiliation{Berkeley Center for Theoretical Physics, University of California, Berkeley, CA 94720, USA\vspace{0.5ex}}
\affiliation{Theoretical Physics Group, Lawrence Berkeley National Laboratory, Berkeley, CA 94720, USA\vspace{0.5ex}}

\author{Lorena Rothen}
\affiliation{Theory Group, Deutsches Elektronen-Synchrotron (DESY), D-22607 Hamburg, Germany\vspace{0.5ex}}

\author{Iain W.~Stewart}
\affiliation{Center for Theoretical Physics, Massachusetts Institute of Technology, Cambridge, MA 02139, USA\vspace{0.5ex}}

\author{Frank J.~Tackmann}
\affiliation{Theory Group, Deutsches Elektronen-Synchrotron (DESY), D-22607 Hamburg, Germany\vspace{0.5ex}}

\author{Hua Xing Zhu}
\affiliation{Center for Theoretical Physics, Massachusetts Institute of Technology, Cambridge, MA 02139, USA\vspace{0.5ex}}
\affiliation{Department of Physics, Zhejiang University, Hangzhou, Zhejiang 310027, China\vspace{0.5ex}}

\date{October 9, 2017}

%%%%%%%%%%%%%%%%%%%%%%%%%%%%%%%%%%%%%%%%%%%%%%%%%%%%%%%%%%%%%%%%%%%%%%%%%%%%%%%%
\begin{abstract}

$N$-jettiness subtractions provide a general approach for performing fully-differential next-to-next-to-leading order (NNLO) calculations. Since they are based on the physical resolution variable $N$-jettiness, $\Tau_N$, subleading power corrections in $\tau=\Tau_N/Q$, with $Q$ a hard interaction scale, can also be systematically computed. We study the structure of power corrections for $0$-jettiness, $\Tau_0$, for the $gg\to H$ process. Using the soft-collinear effective theory we analytically compute the leading power corrections $\alpha_s \tau \ln\tau$ and $\alpha_s^2 \tau \ln^3\tau$ (finding partial agreement with a previous result in the literature), and perform a detailed numerical study of the power corrections in the $gg$, $gq$, and $q\bar q$ channels. This includes a numerical extraction of the $\alpha_s\tau$ and $\alpha_s^2 \tau \ln^2\tau$ corrections, and a study of the dependence on the $\Tau_0$ definition. Including such power suppressed logarithms significantly reduces the size of missing power corrections, and hence improves the numerical efficiency of the subtraction method. Having a more detailed understanding of the power corrections for both $q\bar q$ and $gg$ initiated processes also provides insight into their universality, and hence their behavior in more complicated processes where they have not yet been analytically calculated.

\end{abstract}
%%%%%%%%%%%%%%%%%%%%%%%%%%%%%%%%%%%%%%%%%%%%%%%%%%%%%%%%%%%%%%%%%%%%%%%%%%%%%%%%

\maketitle

%%%%%%%%%%%%%%%%%%%%%%%%%%%%%%%%%%%%%%%%%%%%%%%%%%%%%%%%%%%%%%%%%%%%%%%%%%%%%%%%
\section{Introduction}
\label{sec:intro}
%%%%%%%%%%%%%%%%%%%%%%%%%%%%%%%%%%%%%%%%%%%%%%%%%%%%%%%%%%%%%%%%%%%%%%%%%%%%%%%%

Our ability to perform next-to-next-to-leading order (NNLO) calculations for cross sections of phenomenological importance is crucial for theory predictions to match the precision of Run 2 measurements at the LHC. Due to significant recent progress a number of NNLO subtraction techniques are now available for hadron-hadron collisions, and have been successfully demonstrated both for color-singlet production \cite{Catani:2007vq, Caola:2017dug}, as well as for cross sections involving jets in the final state \cite{GehrmannDeRidder:2005cm,Czakon:2010td,Boughezal:2011jf,Czakon:2014oma,Boughezal:2015aha,Gaunt:2015pea}. However, particularly when final state jets are involved, these techniques remain complicated and computationally expensive. Improving the numerical efficiency and theoretical understanding of NNLO subtraction schemes is therefore of significant interest.

In this paper, we focus on improving the understanding of the infrared structure of $N$-jettiness subtractions \cite{Boughezal:2015aha, Gaunt:2015pea}, which is a nonlocal subtraction method based on the $N$-jettiness resolution variable $\Tau_N$~\cite{Stewart:2009yx, Stewart:2010tn}. $N$-jettiness subtractions provide a powerful and simple method that is in principle applicable for an arbitrary number of jets in the final state. They have been applied to $W/Z/H/\gamma+$ jet at NNLO~\cite{Boughezal:2015dva, Boughezal:2015aha, Boughezal:2015ded, Boughezal:2016dtm,Campbell:2017dqk}, as well as inclusive photon production \cite{Campbell:2016lzl}, and have been implemented in MCFM8 for color-singlet production~\cite{Campbell:2016jau, Campbell:2016yrh, Boughezal:2016wmq, Campbell:2017aul}. They have also been used to calculate single-inclusive jet production in $ep$ collisions at NNLO \cite{Abelof:2016pby}. The $N$-jettiness subtraction scheme has the advantage that it is simple to implement using known NNLO results from the literature, can be interfaced with resummation or parton shower programs,\footnote{Indeed, the first application of $N$-jettiness subtractions was in the \textsc{GENEVA} Monte Carlo \cite{Alioli:2012fc, Alioli:2015toa}.} and is conceptually simple to extend to higher perturbative orders.

An important feature of the $N$-jettiness subtraction scheme is that it is based on a physical infrared-safe observable, $N$-jettiness $\Tau_N$, and the subtraction terms are determined by the behavior of $\Tau_N$ in the soft and collinear limits. Using our understanding of the simplifications of gauge theories in these limits allows the subtraction terms to be systematically computed as an expansion in $\tau \equiv \Tau_N/Q$, with $Q$ a typical hard interaction scale. The leading terms in the $\tau\to 0$ limit are naively nonintegrable divergences that are properly defined as plus-functions, $[\ln^k\tau /\tau]_+$, and are required for the subtractions. These terms are described by well-established factorization formulas valid to all orders in $\alpha_s$. For the case of $N$-jettiness, these formulas were determined in Refs.~\cite{Stewart:2009yx, Stewart:2010tn} using soft collinear effective theory (SCET)~\cite{Bauer:2000ew, Bauer:2000yr, Bauer:2001ct, Bauer:2001yt, Bauer:2002nz}. They are expressed in terms of universal soft, jet, and beam functions. The required ingredients to compute the leading-power subtraction terms at NNLO are the NNLO jet \cite{Becher:2006qw, Becher:2010pd} and beam \cite{Gaunt:2014xga, Gaunt:2014cfa} functions (the spin-dependent quark beam functions were recently computed to NNLO \cite{Boughezal:2017tdd}), which are process independent, as well as the soft function, which depends on the number of external colored partons, $n$, in the Born process. The soft function is known analytically at next-to-leading order (NLO) for arbitrary $n$~\cite{Jouttenus:2011wh}, and at NNLO it is known analytically for $n=2$ \cite{Kelley:2011ng, Monni:2011gb, Hornig:2011iu, Kang:2015moa}, and numerically for $n=3$~\cite{Boughezal:2015eha}, and with a third massive parton \cite{Li:2016tvb}.

The leading-logarithmic (LL) terms at subleading order in $\tau$ were analytically computed at NLO and NNLO for Drell-Yan like processes in Refs.~\cite{Moult:2016fqy,Boughezal:2016zws}. Including these improves the subtractions by an order of magnitude, with further improvements possible by computing additional subleading logarithms. In Ref.~\cite{Moult:2016fqy} it was also shown that the rapidity dependence of the power corrections strongly depends on the observable definition.
For the specific definition of $\Tau_N$ in the hadronic frame that had been used in some implementations, the power corrections grow exponentially with the rapidity $Y$ of the Born system, and the power expansion is in $\Tau_N e^{|Y|}$, instead of $\Tau_N$, causing it to break down at large $Y$. On the other hand, using the definition of $\Tau_N$~\cite{Stewart:2009yx, Stewart:2010tn} that takes into account the boost of the Born system results in a well-behaved power expansion, with power corrections that are approximately flat in $Y$.
Unlike the leading-power factorization, which has the same structure for any color-singlet production, the universality of subleading power corrections is not well understood, and it is therefore important to understand their behavior in other processes.

In this paper, we present a detailed study of the power corrections in $\Tau_0$ for the gluon fusion, $gg\to H$, process. We analytically compute the LL correction at both NLO and NNLO for all partonic channels, namely $gg$, $g q$, and $q\bar q$, and at NLO we also compute the next-to-leading-logarithmic (NLL) contribution for the $q\bar q$ channel, which is the first nonzero contribution from this channel. We then perform a detailed numerical study using $H+1$ jet NLO results from MCFM8~\cite{Campbell:1999ah, Campbell:2010ff, Campbell:2015qma, Boughezal:2016wmq}. This provides both a confirmation of our analytic calculation, and enables us to study the extent to which the power corrections are well described by the LL approximation. We also study the rapidity and observable dependence of the power corrections.

The analytic LL power corrections for $\Tau_0$ in the hadronic frame  for the $gg$ and $gq$ channels were first computed in Ref.~\cite{Boughezal:2016zws}. While we agree with the results of Ref.~\cite{Boughezal:2016zws} for the $gq$ channel, we disagree for the $gg$ channel.\footnote{We also find the same disagreement between our calculation for $q\bar q \to V$ \cite{Moult:2016fqy} and the corresponding $q\bar q\to V$ results given in Ref.~\cite{Boughezal:2016zws} in the $q\bar q$ channel.} We will comment further on this disagreement at the end of \sec{num_results}.  We also note that Ref.~\cite{Boughezal:2016zws} did not perform a numerical study of the partonic channel dependence or of the importance of NLL power-suppressed terms.

The remainder of this paper is organized as follows. In \sec{Njettiness}, we briefly review $N$-jettiness subtractions, focusing in particular on the structure of the power corrections.  In \sec{calc}, we present our analytic calculation of the subleading power logarithms. We perform the calculation both for thrust in $H\to gg$, as well as for beam thrust in $gg\to H$. We also discuss the similarity of the structure of the results with the case of $e^+e^-\to$ dijets and Drell-Yan. In \sec{num_results}, we compare our analytic results with the full nonsingular distribution obtained numerically from MCFM8~\cite{Campbell:1999ah, Campbell:2010ff, Campbell:2015qma, Boughezal:2016wmq}, numerically extract the NLL power corrections, and discuss their importance. In \sec{discuss}, we discuss the importance of the  $\Tau_0$ observable definition, and the dependence of the power corrections on rapidity. We conclude in \sec{conclusions}.

%%%%%%%%%%%%%%%%%%%%%%%%%%%%%%%%%%%%%%%%%%%%%%%%%%%%%%%%%%%%%%%%%%%%%%%%%%%%%%%%
\section{\boldmath $N$-jettiness Subtractions}
\label{sec:Njettiness}
%%%%%%%%%%%%%%%%%%%%%%%%%%%%%%%%%%%%%%%%%%%%%%%%%%%%%%%%%%%%%%%%%%%%%%%%%%%%%%%%

Here we briefly review $N$-jettiness subtractions, focusing on the simplest case of a global subtraction or phase-space slicing,
and using $N = 0$ as relevant for $gg\to H$.
For an in-depth discussion, including a description of how it can be used as a more differential subtraction scheme, see Ref.~\cite{Gaunt:2015pea}.

The cross section $\sigma(X)$, including Born level measurements and cuts $X$, can be expressed as an integral of the differential cross section for $0$-jettiness, $\df\sigma(X)/\df\Tau_0$, as
%%%
\begin{align}
\sigma(X, \Tau_\cut) &\equiv \int^{\Tau_\cut}\!\df\Tau_0\, \frac{\df\sigma(X)}{\df\Tau_0}
\,, \nn \\
\sigma(X)
&= \sigma(X, \Tau_\cut) + \int_{\Tau_\cut}\! \df\Tau_0\, \frac{\df\sigma(X)}{\df\Tau_0}
\,.\end{align}
%%%
For the rest of the paper we will suppress the dependence on $X$. The $N$-jettiness subtraction is implemented by adding and subtracting a subtraction term  $\df\sigma^\sub/\df\Tau_0$, which will ultimately be derived from the singular behavior in the $\Tau_0\to 0$ limit. We can then write the cross section as
%%%
\begin{align} \label{eq:nsub_master}
\sigma
&=
\sigma^\sub(\Tau_{\cut})
+ \int_{\Tau_\cut}\! \df\Tau_0 \frac{\df\sigma}{\df\Tau_0}
+ \bigl[\sigma(\Tau_\cut) - \sigma^\sub(\Tau_\cut) \bigr]
\nn \\
&\equiv \sigma^\sub(\Tau_\cut) + \int_{\Tau_\cut}\! \df\Tau_0 \frac{\df\sigma}{\df\Tau_0}
+ \Delta \sigma(\Tau_\cut)
\,.\end{align}
%%%
In the second term, the restriction $\Tau_0 > \Tau_\cut$ resolves an emission off of the $0$-jet Born configuration, and thus reduces to an NLO calculation of $gg\to H+1$ jet. The last term $\Delta \sigma(\Tau_\cut)$ vanishes by construction as $\Tau_\cut\to 0$ and is eventually neglected.

\begin{figure*}[t]
\includegraphics[width=\columnwidth]{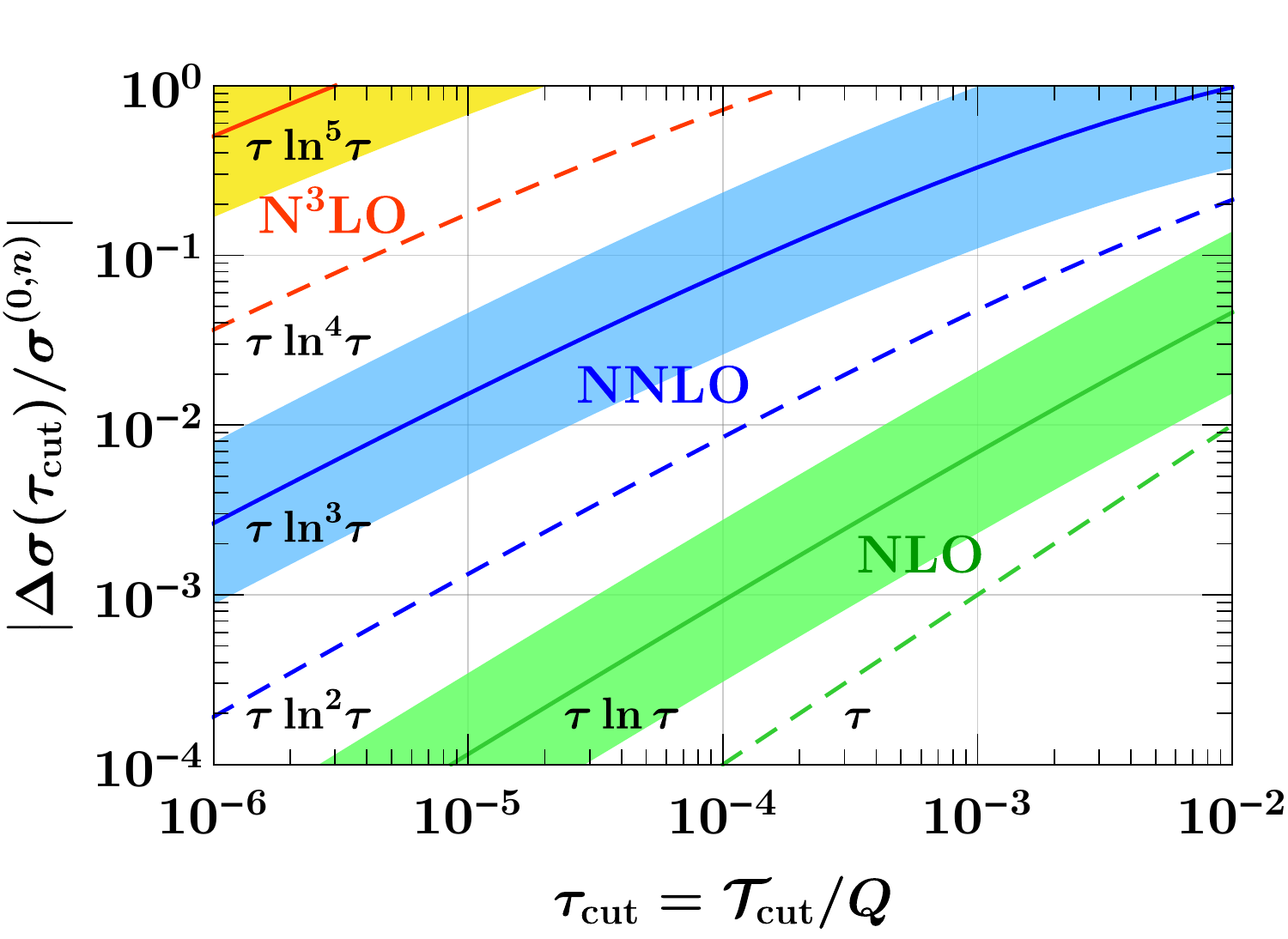}%
\hfill%
\includegraphics[width=\columnwidth]{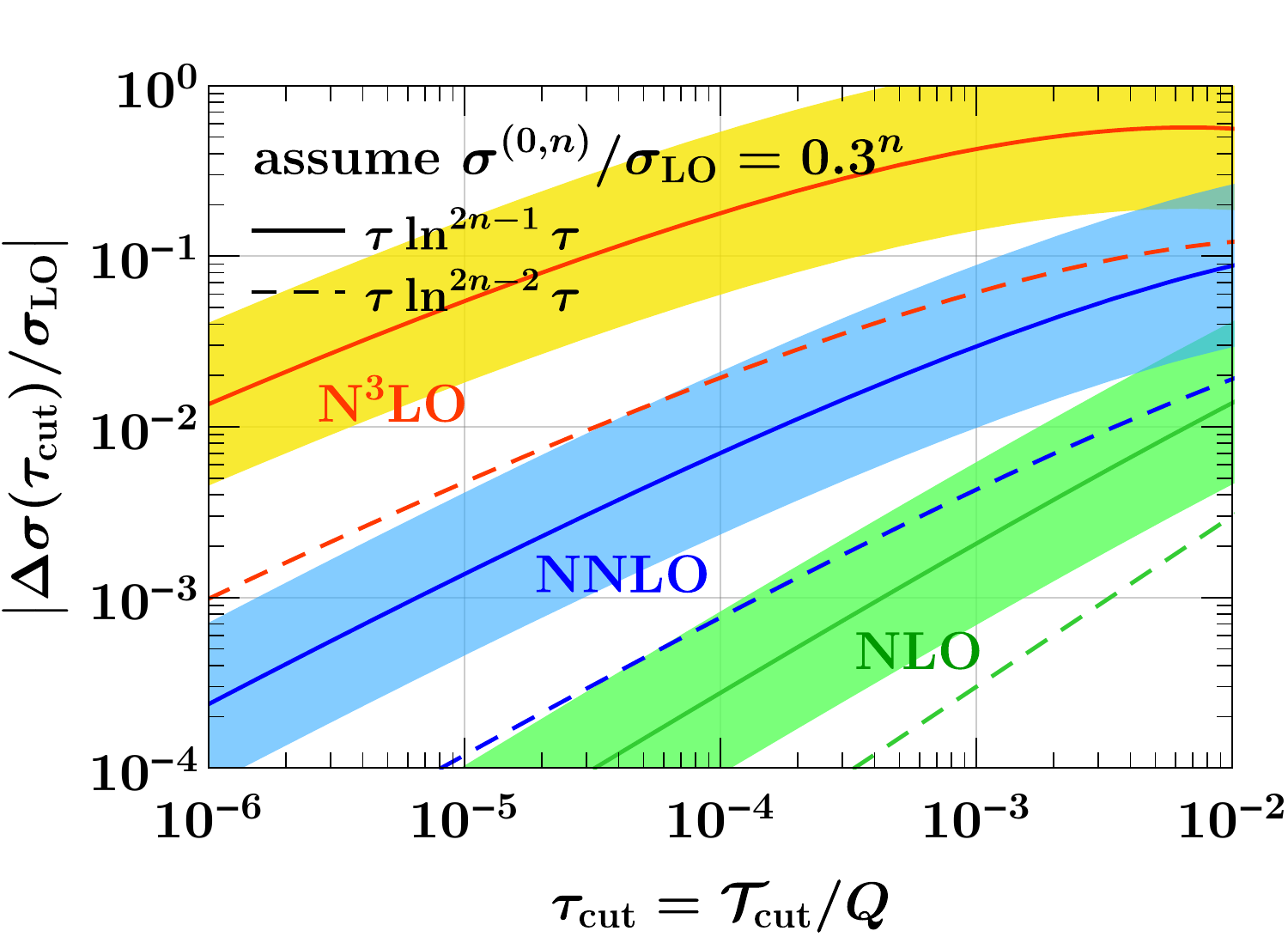}%
\caption{An estimate of the missing power corrections $\Delta \sigma(\tau_\cut)$ based on their functional form at NLO (green), NNLO (blue), and N$^3$LO (orange). The solid (dashed) lines correspond to the (N)LL power corrections, showing the possible improvement by removing the LL corrections. On the left, the estimate is normalized to the full N$^n$LO contribution, while on the right it is normalized to the LO cross section (assuming a 30\% correction of each perturbative order relative to the previous order). In both cases, the bands around the LL estimate illustrates a factor of $3$ variation.}
\label{fig:scaling}
\end{figure*}

We can expand the cross section about the soft and collinear limits in powers of $\tau = \Tau_0/Q \ll 1$. For the case of $gg\to H$, we will take the hard scale to be $Q=m_H$. We write the expanded cross section as
%%%
\begin{align}\label{eq:xsec_expand}
\frac{\df \sigma}{\df\tau}
&= \frac{\df\sigma^{(0)}}{\df\tau} + \frac{\df\sigma^{(2)}}{\df\tau}+ \frac{\df\sigma^{(4)}}{\df\tau} + \dotsb
\,, \\ \nn
\sigma(\tau_\cut)
&= \sigma^{(0)}(\tau_\cut) + \sigma^{(2)}(\tau_\cut) + \sigma^{(4)}(\tau_\cut) + \dotsb
\,.\end{align}
%%%
The superscripts denote the suppression in powers of $\lambda \sim \sqrt{\tau}$. Odd powers of $\sqrt{\tau}$ vanish, as explained in Ref.~\cite{Feige:2017zci}. The leading-power terms $\df\sigma^{(0)}/\df\tau$ and $\sigma^{(0)}(\tau_\cut)$ contain the most singular terms with the scaling
%%%
\begin{align}
\frac{\df\sigma^{(0)}}{\df\tau}
&\sim \delta(\tau)+ \biggl[\frac{ \ord{1} }{\tau}\biggr]_+
\,, \nn \\
\sigma^{(0)}(\tau_\cut) &\sim \ord{1}
\,,\end{align}
%%%
where the $\cO(1)$ in the counting indicates that it is modulo logarithms $\ln^j\tau$ and $\ln^j\tau_\cut$ with $j\geq 1$, respectively. The remaining terms in \eq{xsec_expand},  $\df\sigma^{(2k)}/\df\tau$ and $\sigma^{(2k)}(\tau_\cut)$, contain at most integrable singularities and scale as
%%%
\begin{align}\label{eq:scaling_lam2}
\tau \frac{\df\sigma^{(2k)}}{\df\tau} &\sim \ord{\tau^k}
\,, \quad
\sigma^{(2k)}(\tau_\cut) \sim \ord{\tau_\cut^k}
\,.\end{align}
%%%
The terms with $k = 1$ are the next-to-leading power (NLP) contributions, and are the focus of this paper.

To provide an appropriate subtraction term, $\sigma^\sub(\Tau_\cut)$ must contain all leading-power contributions, but can differ by power-suppressed terms,
%%%
\begin{equation}
\sigma^\sub(\Tau_\cut) = \sigma^{(0)}(\tau_\cut = \Tau_\cut/Q)\, [1 + \ord{\tau_\cut}]
\,.\end{equation}
%%%
Thus, the leading-power terms, obtained from the leading-power $N$-jettiness factorization theorem, provide the minimal subtraction terms for the $N$-jettiness subtraction. In this case, the neglected power corrections $\Delta \sigma(\Tau_\cut)$ are the complete set of power corrections relative to the leading-power factorization.
As pointed out in Ref.~\cite{Gaunt:2015pea}, the inclusion of additional power-suppressed terms in $\sigma^\sub(\Tau_\cut)$ is formally not necessary, as they vanish in the $\tau_\cut\to 0$ limit, but can significantly improve the numerical efficiency of the $N$-jettiness subtractions.

Following Refs.~\cite{Gaunt:2015pea, Moult:2016fqy}, the expected size of the neglected power corrections and the effect of including the dominant power corrections in the subtractions can be easily estimated by their functional form. Writing the perturbative expansion of the cross section in $\alpha_s$ as
%%%
\begin{align}
\frac{\df\sigma^{(k)}}{\df\tau}
&= \sum_{n=0} \frac{\df\sigma^{(k,n)}}{\df\tau} \Bigl( \frac{\alpha_s}{4\pi} \Bigr)^n
\,,\end{align}
%%%
the perturbative structure of the NLP contributions is
%%%
\begin{align}
\tau  \frac{\df\sigma^{(2,n)}}{\df\tau}
&= \tau \sum_{m=0}^{2n-1} \ C^{(2,n)}_m \ln^m \tau
\,,\nn \\
\sigma^{(2,n)}(\tau_\cut)
&=  \tau_\cut \sum_{m=0}^{2n-1}\ A^{(2,n)}_m \ln^{m} \tau_\cut
\,,\end{align}
%%%
where the $A^{(2,n)}_m$ coefficients are directly related to the $C^{(2,n)}_{m'}$ coefficients by integration.
The numerically largest power corrections are then the LL terms at each order, namely the $\alpha_s \,\tau_\cut \ln\tau_\cut$ at NLO and
the $\alpha^2_s\, \tau_\cut \ln^3 \tau_\cut$ at NNLO.
In \fig{scaling} we show an estimate of the neglected power corrections $\Delta\sigma(\tau_\cut)$, based on their logarithmic structure as a function of $\tau_\cut$, relative to the leading-power $\alpha_s^n$ coefficient $\sigma^{(0,n)}$ (on the left) and relative to the LO cross section (on the right). The latter is derived from the former assuming a $30\%$ correction at each order in $\alpha_s$ relative to the contribution from the previous order (corresponding to a $10\%$ correction multiplied by a $C_A$ color factor), and the bands show a factor of three variation of the estimate. The size of the missing terms grows rapidly with the loop order due to the increasing enhancement by additional logarithms. This also highlights that by computing and including the LL power suppressed terms, the numerical efficiency of the $N$-jettiness subtractions can be improved by up to an order of magnitude. We will find that this simple estimate holds for the $gg\to H$ partonic channel, but is not well reproduced for the $gq\to H$ partonic channel, whose LL power correction turns out to be unusually small compared to its NLL power correction.

%%%%%%%%%%%%%%%%%%%%%%%%%%%%%%%%%%%%%%%%%%%%%%%%%%%%%%%%%%%%%%%%%%%%%%%%%%%%%%%%
\section{Calculation}
\label{sec:calc}
%%%%%%%%%%%%%%%%%%%%%%%%%%%%%%%%%%%%%%%%%%%%%%%%%%%%%%%%%%%%%%%%%%%%%%%%%%%%%%%%

In this section, we present our calculation of the subleading power LL coefficients at NLO and NNLO. We perform the calculation for thrust in $H\to gg$ in \subsec{thrust} and for $0$-jettiness (beam thrust) in $gg\to H$ in \subsec{beam_thrust}.  In \sec{compare}, we provide a discussion of the similarities in the analytic structure of the power corrections for $q\bar q$-initiated Drell-Yan and $gg$-initiated $H$ production.

Following our calculation of the fixed-order power corrections for Drell-Yan \cite{Moult:2016fqy}, we use SCET~\cite{Bauer:2000ew, Bauer:2000yr, Bauer:2001ct, Bauer:2001yt, Bauer:2002nz} to systematically organize all sources of power corrections. SCET is an effective field theory for the soft and collinear limits of QCD, which allows for a systematic expansion about the soft and collinear limits in a power counting parameter which for the present case of interest scales as $\lambda^2 \sim \tau$. The effective theory has different quark and gluon fields for soft and collinear particles, which will play an important role in our calculation.

In the effective theory, power corrections arise from three sources, each of which are easy to track. Subleading Lagrangian insertions describe universal corrections to the dynamics of soft and collinear radiation, and are known in the literature to ${\cal O}(\lambda^2)$ \cite{Beneke:2002ph,Chay:2002vy,Manohar:2002fd,Pirjol:2002km,Beneke:2002ni,Bauer:2003mga}. Power-suppressed hard-scattering operators describe subleading-power local corrections to the hard-scattering vertex. Complete operator bases for Drell-Yan and Higgs production to subleading power were derived in Refs.~\cite{Kolodrubetz:2016uim,Feige:2017zci,Moult:2017rpl}. Finally, there are power corrections to the measurement function. These were originally derived for thrust in Ref.~\cite{Freedman:2013vya} using a different formalism than we use here, and in the formalism we use here in Ref.~\cite{Feige:2017zci}.

Following Ref.~\cite{Moult:2016fqy}, we use the scaling of modes in the effective theory to derive general consistency relations allowing for a considerable simplification in the calculation. A general expression for the perturbative expansion of the dimensionally regulated differential cross section at NLP is given by
%%%
\begin{align}\label{eq:constraint_setup}
\frac{\df\sigma^{(2,n)}}{\df\tau}
  = & \sum_{\kappa}\sum_{i=0}^{2n-1} \frac{c_{\kappa,i}}{\epsilon^i} \left( \frac{\mu^{2n}}{Q^{2n} \tau^{m(\kappa)}}   \right)^\epsilon
\nonumber
\\
& + \sum_{\gamma}\sum^{2n-2}_{i=0} \frac{d_{\gamma,i}}{\epsilon^i} 
\left(
\frac{\mu^{2(n-1)}}{Q^{2(n-1)} \tau^{m(\gamma)}}   \right)^\epsilon
\nonumber
\\
& + \dots
\,.\end{align}
%%%
For each contributing loop or real-radiation momentum $k$, we assign a specific label $\kappa,\gamma,\ldots$
for the scalings of the particles, i.e., hard, collinear, or soft, and $m(\kappa)\geq 1$ is an integer. At one loop ($n=1$) there is a single integrated momentum, which can be
%%%
\begin{align} \label{eq:classes1}
\text{soft:} \qquad &\kappa=s\,, \qquad m(\kappa) =2\,,\nn \\
\text{collinear:} \qquad &\kappa=c\,, \qquad m(\kappa)=1 \,.
\end{align}
%%%
At two loops ($n=2$), the possibilities are
%%%
\begin{align} \label{eq:classes2}
\text{hard-collinear:}\qquad  &\kappa = hc\,, \qquad m(\kappa) =1\,,\nn \\
 \text{hard-soft:} \qquad &\kappa= hs\,, \qquad m(\kappa) =2\,,\nn \\
\text{ collinear-collinear:} \qquad & \kappa=cc\,, \qquad m(\kappa) =2\,,\nn \\
 \text{collinear-soft:} \qquad & \kappa= cs\,, \qquad m(\kappa) =3\,,\nn \\
 \text{soft-soft:}  \qquad & \kappa= ss\,,\qquad m(\kappa) =4\,.
 \end{align}
 %%%

The cancellation of $1/\epsilon$ poles, which must occur for an IR-finite observable, implies consistency equations relating the different coefficients. In particular, the power corrections at NLO can be written as~\cite{Moult:2016fqy}
%%%
\begin{align}\label{eq:constraints_NLO}
\frac{\df\sigma^{(2,1)}}{\df\tau}
&= c_{c,1} \ln \tau
+\text{const}
\,,\end{align}
%%%
and at NNLO as~\cite{Moult:2016fqy}
%%%
\begin{align}\label{eq:constraints_final}
\frac{\df\sigma^{(2,2)}}{\df\tau}
&= c_{hc,3} \ln^3 \tau
+ (c_{hc,2} + c_{ss,2} + d_{c,2}) \ln^2 \tau
\nn \\ & \quad
+ \left(-c_{cs, 1} + c_{hc, 1} - 2 c_{ss, 1} + d_{c, 1}    \right) \ln\tau
\nn\\ & \quad
+  d_{c, 2} \ln\frac{Q^2}{\mu^2} \ln\tau
+\text{const}
\,.\end{align}
%%%
Writing the LL contribution purely in terms of the collinear or hard-collinear coefficient significantly simplifies the calculation, since we only need to consider a two particle collinear phase space to compute the leading logarithm. The one-loop hard matching can be extracted from the amplitudes for $H\to 3$ partons, which are known to NLO \cite{Schmidt:1997wr} and NNLO \cite{Gehrmann:2011aa}.

%===============================================================================
\subsection{\boldmath $2$-jettiness in $H \to$ gg}
\label{subsec:thrust}
%===============================================================================

We begin by computing $2$-jettiness in $H\to gg$, which for massless partons is equivalent to thrust \cite{Farhi:1977sg}, for which the exact one-loop result can easily be computed and will provide a cross check on our results.%
\footnote{The full NLO result for thrust in $H\to gg$ was also computed recently in Ref.~\cite{Mo:2017gzp} in a different context.}
The thrust measurement function is defined by
%%%%
\begin{align}\label{eq:thrust_defn}
\tau=1- \text{max}_{\hat t} \frac{\sum_i |\hat t \cdot \vec p_i|}{\sum_i |\vec p_i|}\,.
\end{align}
%%%%
We focus on the $\alpha_s \ln\tau$ and $\alpha^2_s \ln^3 \tau$ terms in $\df\sigma^{(2,n)}$.
We will discuss in some detail the structure of the calculation at NLO, focusing on the different types of power corrections,
and the cancellation of $1/\epsilon$ poles.
We then use \eq{constraints_final} to extend this calculation to NNLO.

%~~~~~~~~~~~~~~~~~~~~~~~~~~~~~~~~~~~~~~~~~~~~~~~~~~~~~~~~~~~~~~~~~~~~~~~~~~~~~~~
\subsubsection{$H\to gg$ Power Corrections at NLO}
%~~~~~~~~~~~~~~~~~~~~~~~~~~~~~~~~~~~~~~~~~~~~~~~~~~~~~~~~~~~~~~~~~~~~~~~~~~~~~~~

\begin{figure}[t]
\subfigure[]{\includegraphics[width=0.49\columnwidth]{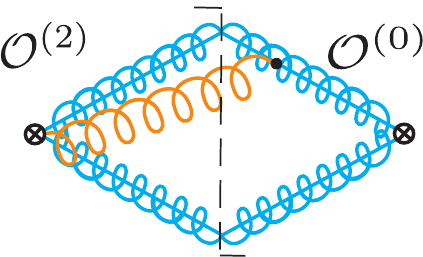}}
\hfill
\subfigure[]{\includegraphics[width=0.47\columnwidth]{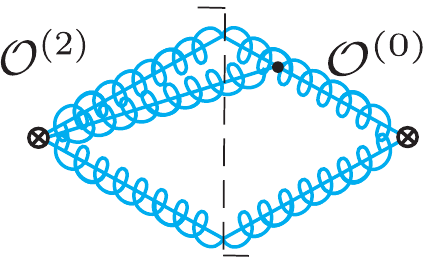}}
\caption{Representative NLO diagrams for Category 1.  In (a) a gluon becomes soft, and in (b) two gluons become collinear. Collinear particles are shown in light blue, soft particles in orange. The power counting of the hard-scattering operators and Lagrangian insertions is explicitly indicated.}
\label{fig:category1}
\end{figure}

Studying contributions from the complete basis of SCET operators \cite{Moult:2017rpl} and Lagrangian insertions, there are four contributions to the leading logarithm at NLO. As for the case of Drell-Yan~\cite{Moult:2016fqy}, we can group these into two categories, each of which separately exhibits the cancellation of $1/\eps$ poles:
\begin{itemize}
\item Category 1: Purely gluonic contributions, where two gluons become collinear, or a gluon becomes soft.
\item Category 2: Contributions involving quarks, where either two quarks become collinear, or a quark becomes soft.
\end{itemize}
The leading logarithm at leading power only comes from the purely gluonic contribution, to which Category 1 contains the NLP corrections. On the other hand, Category 2 has no leading-power analogue and gives rise to new color structures at subleading power. The behavior of the two categories is very similar to the corresponding categories for Drell-Yan, and we will highlight these similarities in \sec{compare}.

Since Category 1 has the same partonic content as the leading-power contribution, it has three possible sources of power corrections: corrections from Lagrangian insertions, corrections from subleading hard-scattering operators, and corrections from the phase space or measurement. The required subleading hard-scattering operators, which involve both additional soft or collinear fields were given in Ref.~\cite{Moult:2017rpl}. Representative diagrams are shown in \fig{category1}, and are of the form of an $\cO(\lambda^2)$ contribution interfered with the leading-power contribution. This form is guaranteed by the  Low-Burnett-Kroll theorem \cite{Low:1958sn,Burnett:1967km}. Subleading power corrections to the thrust measurement function were derived in Refs.~\cite{Freedman:2013vya,Feige:2017zci}. They are found to not give a LL contribution. On the other hand, corrections to the phase space give rise to LL contributions.

\begin{figure}[t]
\subfigure[]{\includegraphics[width=0.49\columnwidth]{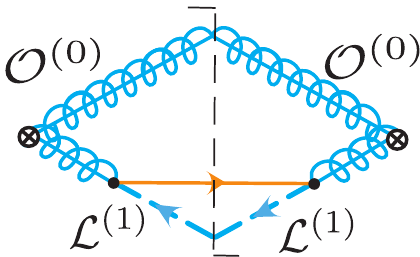}}
\hfill
\subfigure[]{\includegraphics[width=0.49\columnwidth]{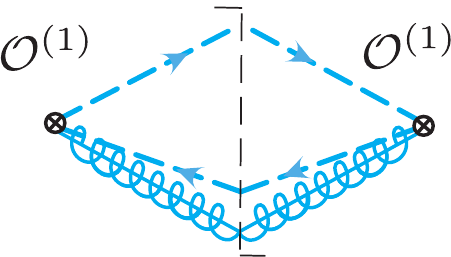}}
%%%
\caption{Representative NLO diagrams for Category 2. In (a) a quark becomes soft, and in (b) a quark and a gluon become collinear. The power counting of the hard-scattering operators and Lagrangian insertions is explicitly indicated.}
\label{fig:category2}
\end{figure}

Summing the different contributions, we find for Category 1
%%%
\begin{align}\label{eq:cat1_NLO}
\frac{1}{\sigma_0}\frac{\df\sigma_\mathrm{Cat.1}^{(2,1)}}{\df\tau}
&= 8C_A \biggl[ \biggl( \frac{1}{\epsilon}+\ln\frac{\mu^2}{Q^2 \tau} \biggr)
- \biggl( \frac{1}{\epsilon} +\ln\frac{\mu^2}{Q^2 \tau^2} \biggr) \biggr]
\nn \\
&= 8C_A \ln \tau
\,.\end{align}
%%%
Here we explicitly see the cancellation of the $1/\epsilon$ poles between the soft and collinear diagrams, which are separated in the first equality.

The contributions to Category 2 obtain their power suppression either from subleading Lagrangian insertions or from subleading hard-scattering operators. Since they have no leading-power analogue, we do not need to include power corrections to the phase space or measurement, as these would be additionally power suppressed. Two contributing diagrams are shown in \fig{category2}. In  \fig{category2} a) we illustrate the contribution from a soft quark, which is described by a subleading Lagrangian insertion, and in \fig{category2} b) we illustrate the collinear limit between a quark and a gluon, which arises from a subleading hard-scattering operator.

Summing the different contributions for Category 2, we find
%%%
\begin{align}
\frac{1}{\sigma_0}\frac{\df\sigma_\mathrm{Cat.2}^{(2,1)}}{\df\tau}
&= 8n_f T_F \biggl[ -\biggl(\frac{1}{\epsilon} + \ln\frac{\mu^2}{Q^2 \tau} \biggr)
+ \biggl(\frac{1}{\epsilon} + \ln\frac{\mu^2}{Q^2 \tau^2}  \biggr) \biggr]
\nn \\
&= -8n_f T_F \ln\tau
\,.\end{align}
%%%
Here $n_f$ denotes the number of light flavors, and $T_F=1/2$. In the first equality, we have separated the contributions from the soft and collinear diagrams, whose $1/\eps$ poles cancel.

The final result for this term in the cross section is
%%%
\begin{equation}
\frac{1}{\sigma_0}\frac{\df\sigma^{(2,1)}}{\df\tau}
= (8C_A-8n_f T_F) \ln\tau
\,.\end{equation}
%%%
This result can be explicitly checked by computing the exact NLO result for thrust in $H\to gg$ and expanding in the $\tau\to0$ limit.

%~~~~~~~~~~~~~~~~~~~~~~~~~~~~~~~~~~~~~~~~~~~~~~~~~~~~~~~~~~~~~~~~~~~~~~~~~~~~~~~
\subsubsection{$H\to gg$ Power Corrections at NNLO}
%~~~~~~~~~~~~~~~~~~~~~~~~~~~~~~~~~~~~~~~~~~~~~~~~~~~~~~~~~~~~~~~~~~~~~~~~~~~~~~~

\begin{figure}[t]
\subfigure[]{\includegraphics[width=0.49\columnwidth]{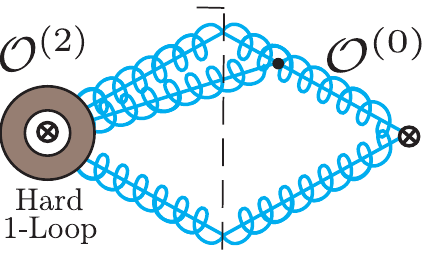}}
\hfill
\subfigure[]{\includegraphics[width=0.49\columnwidth]{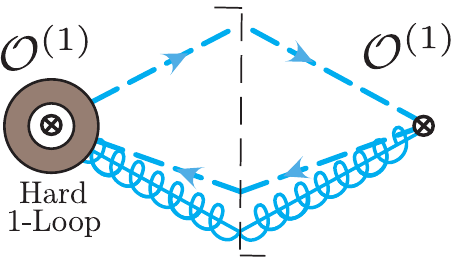}}
%%%
\caption{Two-loop hard-collinear contributions, which are used to compute the LL divergence at subleading power. The grey circle represents a one-loop hard virtual correction. (a) shows the contributions to category 1, when two gluons become collinear, and (b) shows the contributions to category 2 (b), when a quark and a gluon become collinear.}
\label{fig:hc2loop}
\end{figure}

Using the consistency relation of \eq{constraints_final} it is straightforward to extend the NLO result to NNLO. The LL term at NNLO can be computed from the one-loop hard corrections to the collinear contributions at NLO, as shown in \eq{constraints_final}, and illustrated in \fig{hc2loop}. The amplitudes for $H\to 3$ partons are known to NNLO \cite{Gehrmann:2011aa}. From these, we find
%%%
\begin{align}
\frac{1}{\sigma_0}\frac{\df\sigma^{(2,2)}}{\df\tau}
&= \Bigl[ -32 C_A^2 +16n_fT_F(C_F + C_A) \Bigr] \ln^3\tau
\label{eq:thrustnnlo}
\,,\end{align}
%%%
where the two terms arise from Category 1 and 2, respectively. The color structure for the Category 2 result is unusual for a LL contribution.

%===============================================================================
\subsection{\boldmath $0$-jettiness in $gg\to H$}
\label{subsec:beam_thrust}
%===============================================================================

Having understood the analytic calculation of the LL power corrections for 2-jettiness in $H\to gg$, we now turn to $0$-jettiness in $gg\to H$. Here, the additional complications are the parton distribution functions (PDFs), and in particular the definition of the $0$-jettiness measure, which has been discussed in Ref.~\cite{Moult:2016fqy}.

We begin by defining our conventions for the kinematics. We define $q^\mu$, $Q$, and $Y$ as the total momentum, invariant mass, and rapidity of the color-singlet system,
%%%
\begin{equation}
Q = \sqrt{ q^2 }
\,,\qquad
Y = \frac{1}{2} \ln\frac{q^-}{q+}
\,,\end{equation}
%%%
and take the incoming partonic momenta to be
%%%
\begin{align}
p_a^\mu &= x_a \Ecm \frac{n^\mu}{2}
\,,\qquad
x_a \Ecm = Q e^Y
\,, \nn \\
p_b^\mu &= x_b \Ecm \frac{\bn^\mu}{2}
\,,\qquad
x_b \Ecm = Q e^{-Y}
\,,\end{align}
%%%
where $n^\mu = (1, \hat z)$, $\bn^\mu = (1, -\hat z)$, and $\hat z$ is the beam axis.

In Ref.~\cite{Moult:2016fqy} it was found that the structure of the power corrections depends strongly on the choice of the $0$-jettiness measure. In particular, it was shown that the definition of $0$-jettiness that takes into account the boost of the color-singlet Born system (in this case the Higgs), which we will refer to as the leptonic definition, has a well-behaved power expansion, while $0$-jettiness defined in the hadronic center-of-mass frame has a poorly-behaved power expansion with power corrections growing exponentially with $Y$.

We define the dimensionful and dimensionless versions of $0$-jettiness as
%%%
\begin{align}
\Tau_0^x
&= \sum_k \min \Bigl\{ \lambda_x \,p_k^+, \lambda_x^{-1}\,p_k^- \Bigr\}
\,, \qquad
\tau^x \equiv \frac{\Tau_0^x}{Q}
\,,\end{align}
%%%
with the measures for the different definitions given by
%%%
\begin{align}\label{eq:Njet_def}
\text{leptonic:} && \lambda &= \sqrt{\frac{q^-}{q^+}} = e^Y\,, \qquad \tau=\frac{\Tau_0}{Q}
\,, \nn \\
\text{hadronic:} && \lambda_\hadcm &= 1\,, \qquad \tau^{\hadcm}=\frac{\Tau_0^\hadcm}{Q}
\,.\end{align}
%%%
In both cases, the sum runs over all particles in the final state excluding the Higgs, and the momenta $p_k$ are defined in the hadronic center-of-mass frame.

In this section, we derive analytic results for the power corrections with the leptonic definition. The well-behaved power expansion for the leptonic definition, allows for a precise comparison of our analytic results with the numerical results for the power corrections extracted from the full $H+1$ jet numerical NLO calculation, as will be discussed in \sec{num_results}. The analytic results for the hadronic definition are given in \sec{discuss}.

The hadronic cross section is written as a convolution of the PDFs, $f_i$, and the partonic cross sections, $\df\hat\sigma_{ij}(\xi_a,\xi_b)$, as
%%%
\begin{equation}
\df \sigma
= \sum_{ij}\int\! \df \xi_a \df \xi_b\, f_i(\xi_a)\, f_j(\xi_b)\,\df\hat\sigma_{ij}(\xi_a,\xi_b)
\,.\end{equation}
%%%
The leading-order partonic cross section, to which we will normalize, is given by 
%%%
\begin{equation} \label{eq:sigmapartLO}
\frac{\df\hat\sigma_{gg}^{(0,0)}(\xi_a, \xi_b; X)}{\df Q^2\, \df Y\,\df \tau}
= \sigma_0(Q,X)\,\delta_a \delta_b \, \delta(\tau)
\,,\end{equation}
%%%
where
%%%
\begin{equation} \label{eq:deltaab}
\delta_a \equiv \delta(\xi_a - x_a)
\,, \qquad
\delta_b \equiv \delta(\xi_b - x_b)
\,,\end{equation}
%%%
and $\sigma_0(Q,X)$ is the $gg\to H$ Born cross section, including any cuts, $X$, on the Born phase space.

At subleading power, derivatives of the PDFs enter due to the routing of small momentum components into the incoming collinear lines. Explicitly, these will appear in the form
%%%
\begin{equation}
f_i\biggl[\xi\Bigl(1 + \frac{k}{Q}\Bigr)\biggr] = f_i(\xi) + \frac{k}{Q}\, \xi f_i'(\xi) + \dotsb
\,,\end{equation}
%%%
where $k/Q\sim \tau$.

%~~~~~~~~~~~~~~~~~~~~~~~~~~~~~~~~~~~~~~~~~~~~~~~~~~~~~~~~~~~~~~~~~~~~~~~~~~~~~~~
\subsubsection{Results for $gg\to H$}
%~~~~~~~~~~~~~~~~~~~~~~~~~~~~~~~~~~~~~~~~~~~~~~~~~~~~~~~~~~~~~~~~~~~~~~~~~~~~~~~

\begin{figure}[t]
\subfigure[]{\includegraphics[width=0.42\columnwidth]{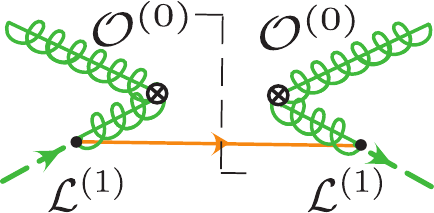}}
\hfill
\subfigure[]{\includegraphics[width=0.51\columnwidth]{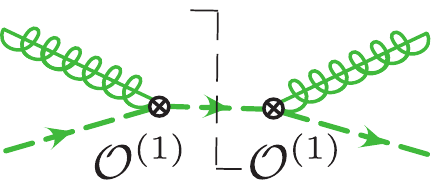}}
%%%
\caption{Representative diagrams contributing to $\sigma_{gq}$, where either a soft quark crosses the cut (a), or a collinear quark crosses the cut (b). One-loop corrections to these diagrams give rise to the $C_F(C_F+C_A) \ln^3\tau$ structure at NNLO.}
\label{fig:beamthrust}
\end{figure}

We now give the NLO and NNLO results for the LL contribution at NLP with the leptonic definition, $\tau=\Tau_0/Q$. We define our notation for the NLP partonic cross section as
%%%
\begin{align} \label{eq:NLP}
\frac{\df\hat\sigma^{(2,n)}_{ij}(\xi_a, \xi_b; X)}{\df Q^2\,\df Y\,\df\tau}
= \sigma_0(Q,X) \sum_{m=0}^{2n-1} C^{(2,n)}_{ij,m}(\xi_a, \xi_b)\ln^m\tau
\,,\end{align}
%%%
where $\sigma_0(Q,X)$ is the gluon-fusion Born cross section defined via \eq{sigmapartLO}.

\begin{figure}[t]
\includegraphics[width=0.45\columnwidth]{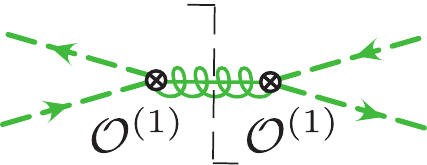}
\caption{Diagram contributing to $\sigma_{q\bar q}$, arising from a subleading operator where a collinear gluon crosses the cut. There is no corresponding soft diagram. Unlike for the other channels, this channel is IR finite at lowest order, so there is no constraint from the cancellation of soft and collinear IR poles.}
\label{fig:beamthrust_qq}
\end{figure}

Crossing the diagrams of \figs{category1}{category2}, there are three distinct partonic channels at NLO: $\hat \sigma_{gg}$,
$\hat \sigma_{gq}$ (trivially related to $\hat \sigma_{g\bar q}$, $\hat \sigma_{qg}$, $\hat \sigma_{\bar q g}$) and $\hat \sigma_{q\bar q}=\hat \sigma_{\bar q q}$. At NNLO we additionally have $\hat \sigma_{q q'}$, where $q'$ is a quark or antiquark of unrelated flavor to $q$. For notational simplicity, at NNLO we will take $\hat \sigma_{q q'}$ to include $\hat \sigma_{q\bar q}$. Representative diagrams for the $gq$ channel involving soft and collinear emissions are shown in \fig{beamthrust}. Unlike the $gg$ and $gq$ channels, the $q\bar q$ channel does not contribute a leading logarithm. At $\cO(\alpha_s)$, this implies that it is a constant, while at $\cO(\alpha_s^2)$ it contributes a $\ln^2\tau$. At NLO it has only a collinear contribution, which is shown in \fig{beamthrust_qq}, and is unconstrained by the consistency relations from the cancellation of $1/\epsilon$ poles.
In addition to not contributing a leading logarithm, the $q\bar q$ channel is also suppressed by the parton luminosities, and is therefore numerically irrelevant for controlling the $N$-jettiness power corrections for the case of $gg\to H$.
Nevertheless, it provides insight into the structure of the power corrections at NLL in the simplest possible context, namely when there are no LL power corrections, and we therefore compute it at NLO.

To compute the LL coefficients $C^{(2,1)}_{ij,1}$ and $C^{(2,2)}_{ij,3}$, we cross our results for thrust computed in \subsec{thrust}. Taking into account the modified definition of the measurement function as well as the corrections from PDFs, we find for the NLO coefficients
%%%
\begin{align} \label{eq:NLOResult}
C_{gg, 1}^{(2,1)}(\xi_a, \xi_b)
&=  8C_A\left(\delta_a \delta_b+\frac{\delta_a' \delta_b}{2}
+\frac{\delta_a \delta_b'}{2}\right)
\,, \nn \\
C_{gq, 1}^{(2,1)}(\xi_a, \xi_b)
&= -2 C_F \,\delta_a \delta_b
\,, \nn \\
C_{q\bar q, 1}^{(2,1)}(\xi_a, \xi_b)
&= 0
\,.\end{align}
%%%
The derivatives of the delta functions are defined as
%%%
\begin{align}
\delta'_a \equiv x_a\, \delta'(\xi_a - x_a)
\,, \qquad
\delta'_b \equiv x_b\, \delta'(\xi_b - x_b)
\,,\end{align}
%%%
which translate into the above-mentioned PDF derivatives in the hadronic cross section.
They only appear in the $gg$ coefficients, because the $gq$ coefficient has no analog at leading power that is sufficiently singular.
Repeating the analysis at NNLO, we obtain for the NNLO coefficients
%%%
\begin{align} \label{eq:NNLOResult}
C_{gg, 3}^{(2,2)}(\xi_a, \xi_b)
&= -32 C_A^2 \left(\delta_a \delta_b+\frac{\delta_a' \delta_b}{2}
+\frac{\delta_a \delta_b'}{2}\right)
\,, \nn \\
C_{gq, 3}^{(2,2)}(\xi_a, \xi_b)
&=  4 C_F(C_F+C_A)\, \delta_a \delta_b
\,, \nn \\
C_{q q', 3}^{(2,2)}(\xi_a, \xi_b)
&=  0
\,. \end{align}
%%%
Our results for $C_{gq, 1}^{(2,1)}$ and $C_{gq, 3}^{(2,2)}$ agree with those in Ref.~\cite{Boughezal:2016zws}. We discuss the difference between our results in the $gg$ channel and those of Ref.~\cite{Boughezal:2016zws} at the end of \sec{num_results}.

For the $q\bar q$ channel, the NLL coefficient at NLO arising from the diagram in \fig{beamthrust_qq} is given by
%%%
\begin{align} \label{eq:NLOResult_qq}
C_{q\bar q, 0}^{(2,1)}(\xi_a, \xi_b)
&= 16 \frac{C_F}{N_c} \biggl[ \frac{\delta_b}{x_a}\Bigl( 1- \frac{x_a}{\xi_a}\Bigr)^2 + (a\leftrightarrow b) \biggr]
\,.
\end{align}
%%%
We expect this to be representative of the typical structures that will appear for NLL power corrections, namely kernels of $x_{a,b},~\xi_{a,b}$, much like at leading power. It would be interesting to also calculate the NLL power corrections for the other partonic channels, but this is beyond the scope of this paper.

%~~~~~~~~~~~~~~~~~~~~~~~~~~~~~~~~~~~~~~~~~~~~~~~~~~~~~~~~~~~~~~~~~~~~~~~~~~~~~~~
\subsection{Comparison with Drell-Yan}\label{sec:compare}
%~~~~~~~~~~~~~~~~~~~~~~~~~~~~~~~~~~~~~~~~~~~~~~~~~~~~~~~~~~~~~~~~~~~~~~~~~~~~~~~

It is interesting to compare the structure of the calculation performed in this section with our results for Drell-Yan presented in Ref.~\cite{Moult:2016fqy}. In both cases, there are two categories of terms in the organization of the effective theory, each with a hard-scattering operator involving an additional collinear field, and a contribution involving a soft parton. In both cases, the two categories are distinguished by whether the hard-scattering operators involving an additional collinear field have the same number of quarks in each collinear sector as the corresponding leading-power operator. In the Drell-Yan case, there is an operator involving collinear quarks in opposite collinear sectors, with an additional collinear gluon in one sector, as well as an operator involving two collinear quarks in the same sector recoiling against a collinear gluon in the opposite sector. This is nearly identical to the structure of the operators for $gg\to H$. Furthermore, in both cases one has a contribution from the soft quark Lagrangian and from a soft gluon.

One other interesting aspect of this organization is that the $\cO(\lambda^2)$ contribution associated with category 1 with a soft gluon, appears as the interference of an $\cO(\lambda^2)$ operator with the leading-power operator. For the soft contribution, this is guaranteed by the Low-Burnett-Kroll theorem \cite{Low:1958sn,Burnett:1967km}, although it is not obvious why it also appears in this form for the collinear sector. While the consistency relations relate the coefficients of the poles, they could in principle be satisfied with multiple collinear contributions. Ultimately the derivation of the full renormalization group consistency equation should shed some light on this. On the other hand, the contributions involving the soft quark Lagrangian, or the hard-scattering operators with different numbers of quarks in each collinear sector than at leading power always contribute to the cross section by multiplying their hermitian conjugate as $\cO(\lambda) \times \cO(\lambda)$.

It is also interesting to compare the structure of the radiative corrections in the two cases. In particular, we can compare the ratios of the one-loop and two-loop coefficients
for each of the different categories. We find
\begin{align}
\frac{\df\sigma_\mathrm{Cat.1}^{(2,2),\text{Higgs}}}{\df\sigma_\mathrm{Cat.1}^{(2,1),\text{Higgs}}} &=-4C_A\,, \qquad \frac{\df\sigma_\mathrm{Cat.1}^{(2,2),\text{Drell-Yan}}}{\df\sigma_\mathrm{Cat.1}^{(2,1),\text{Drell-Yan}}}  =-4C_F\,, \nn \\
\frac{\df\sigma_\mathrm{Cat.2}^{(2,2),\text{Higgs}}}{\df\sigma_\mathrm{Cat.2}^{(2,1),\text{Higgs}}} &=-2(C_A+C_F)=\frac{\df\sigma_\mathrm{Cat.2}^{(2,2),\text{Drell-Yan}}}{\df\sigma_\mathrm{Cat.2}^{(2,1),\text{Drell-Yan}}} \,.
\end{align} 
Interestingly, these are identical up to the exchange $C_A\leftrightarrow C_F$. This is of course true for the leading logarithms at leading power.  In both cases for Category 1, which behaves like at leading power, the scaling appears to be as arising from the cusp anomalous dimension. Ideally this result could be derived to all orders by studying the renormalization group evolution of the operators. For the other channel, the scaling in both cases is identical, and is a linear combination of the color Casimirs. It would be interesting to understand its all-orders structure, in particular whether it arises from a linear combination of two cusp anomalous dimensions, and if so, why this combination is identical for $gg$ and $q\bar q$ initiated processes. It would also be interesting to understand to what extent similar relations persist for the subleading logarithms, as well as at higher orders in $\alpha_s$, or higher powers.

Recently, Ref.~\cite{DelDuca:2017twk} appeared, which proves the universality of power corrections in the threshold limit at NLO. It does not, however, discuss contributions involving soft quarks, the extension away from the threshold limit, or beyond NLO. These are all directions that would be interesting to pursue.

%~~~~~~~~~~~~~~~~~~~~~~~~~~~~~~~~~~~~~~~~~~~~~~~~~~~~~~~~~~~~~~~~~~~~~~~~~~~~~~~
\section{Numerical results}\label{sec:num_results}
%~~~~~~~~~~~~~~~~~~~~~~~~~~~~~~~~~~~~~~~~~~~~~~~~~~~~~~~~~~~~~~~~~~~~~~~~~~~~~~~

\begin{figure*}[t!]
\includegraphics[width=\columnwidth]{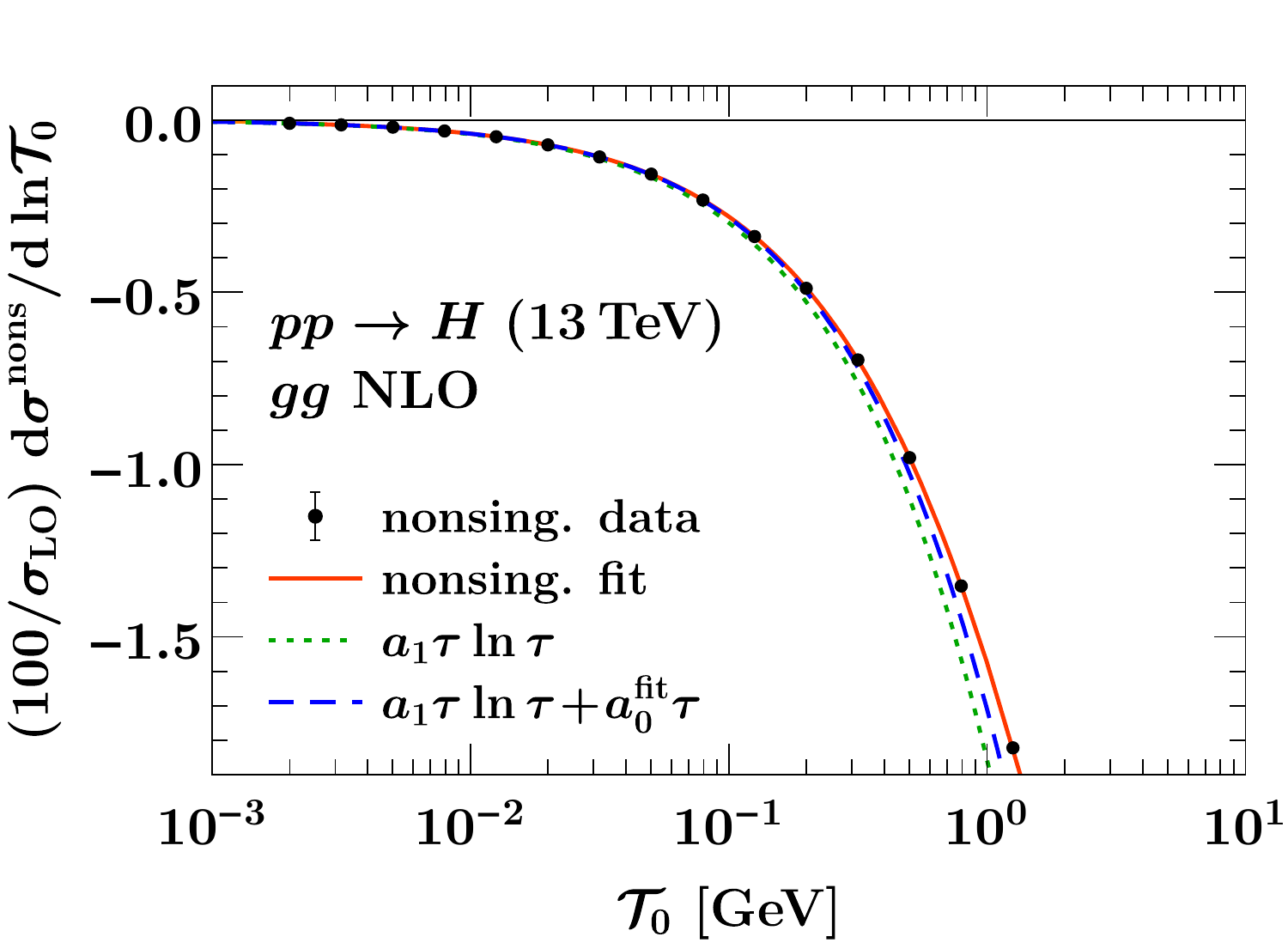}%
\hfill
\includegraphics[width=\columnwidth]{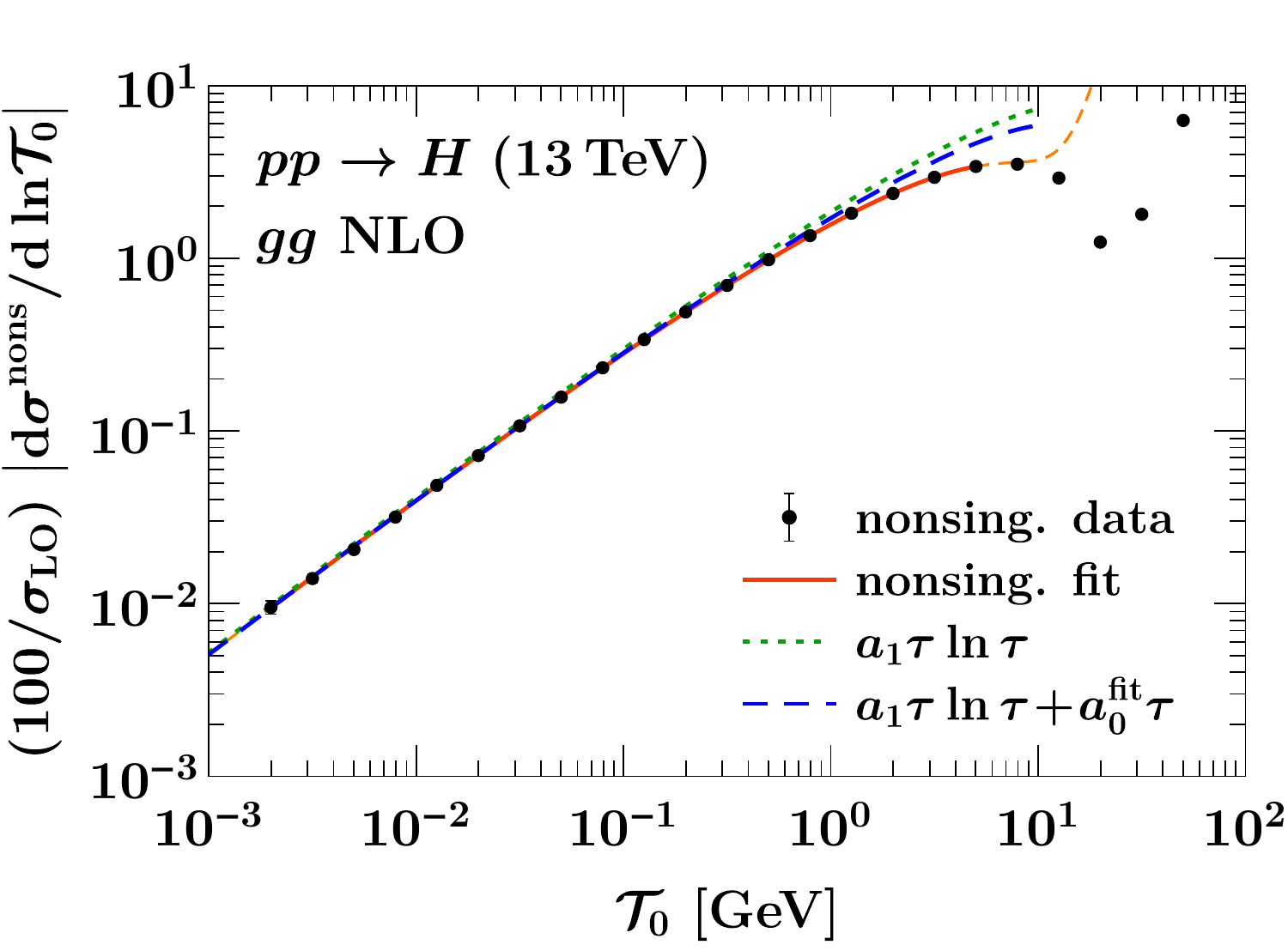}%
\\
\includegraphics[width=\columnwidth]{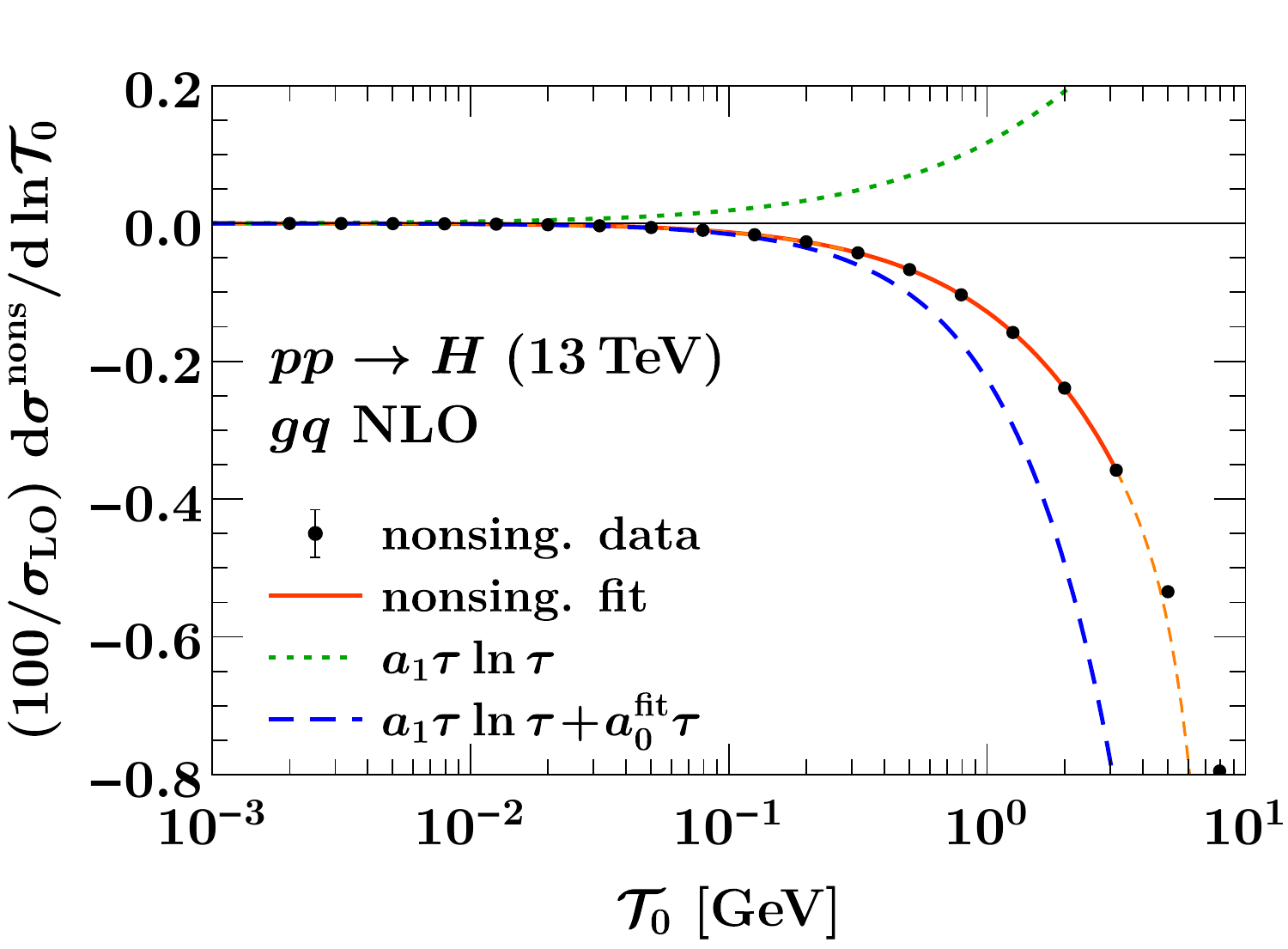}%
\hfill
\includegraphics[width=\columnwidth]{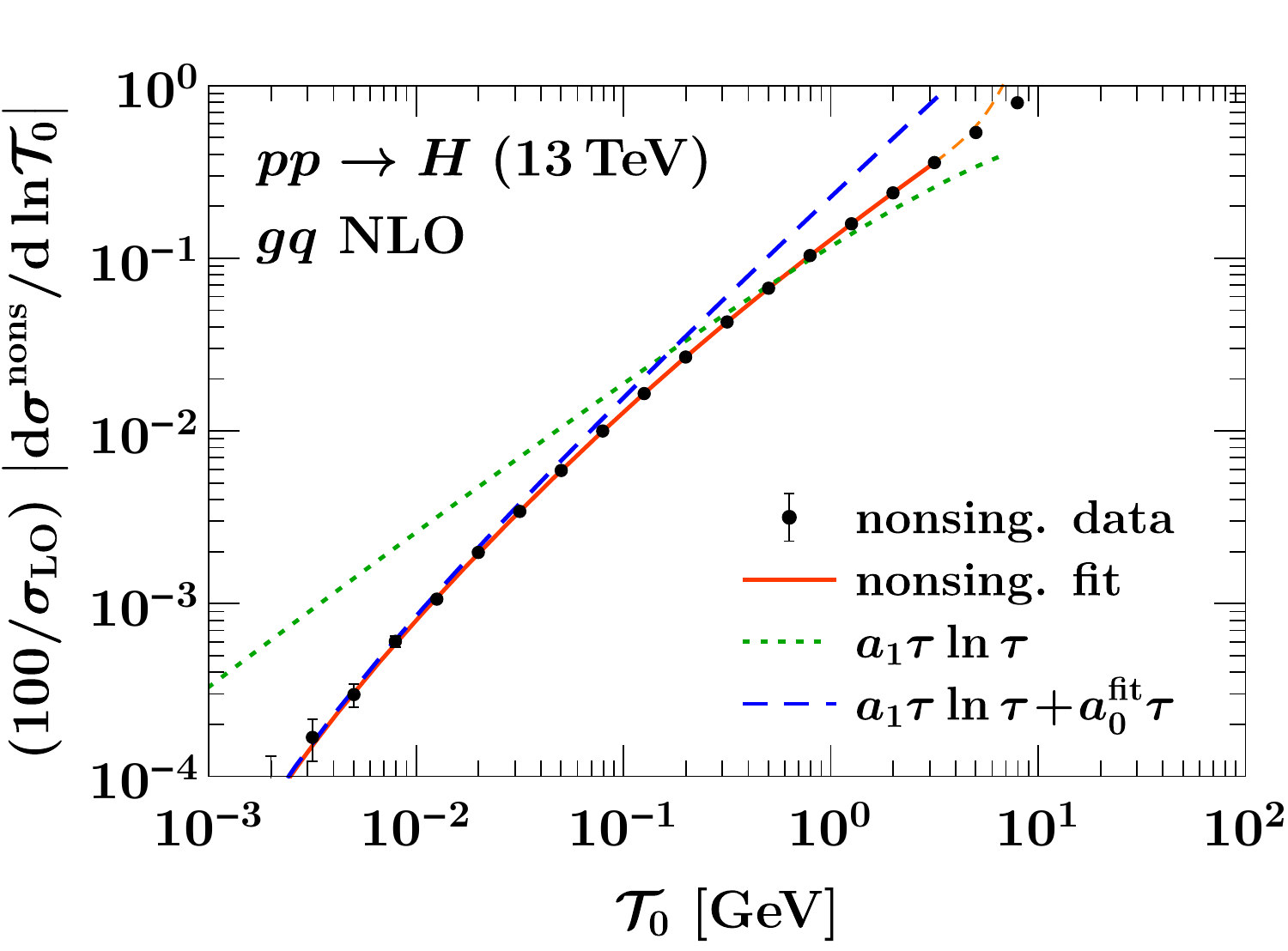}%
%%%
\caption{Fit to the $\ord{\alpha_s}$ nonsingular corrections for beam thrust in the $gg$ channel (top row) and the $gq$ channel (bottom row). The fit functions are defined in \eq{fitfun} and yield the solid red line, while the dashed blue line shows the result when only the leading coefficient from the red solid fit is retained. (The light dashed orange line shows an extrapolation of the fit result beyond the fit region.) The plots in the right column are identical to those in the left column, but show the absolute value on a logarithmic scale.}
\label{fig:fitNLO}
\end{figure*}

\begin{figure*}[t!]
\includegraphics[width=\columnwidth]{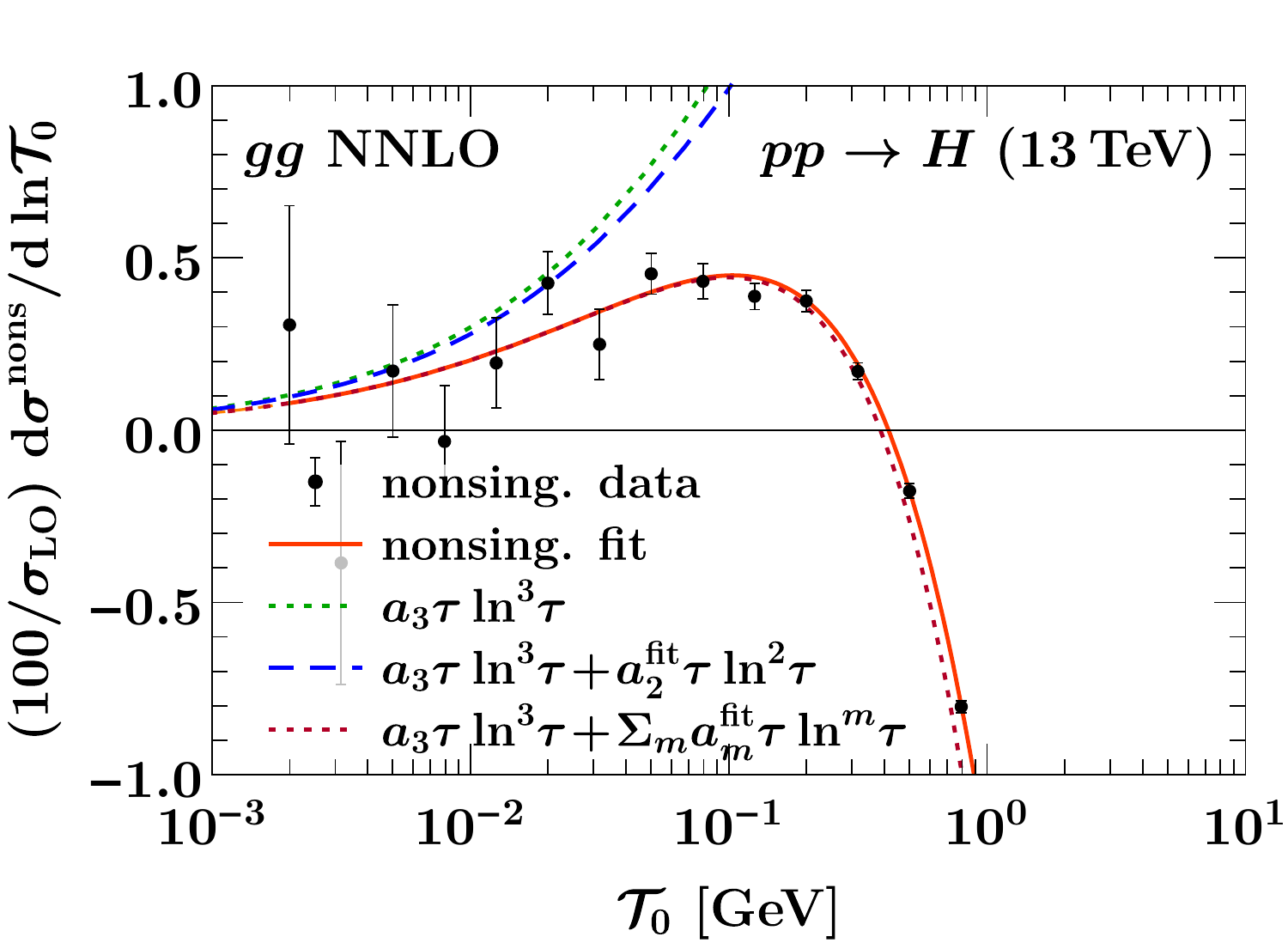}%
\hfill
\includegraphics[width=\columnwidth]{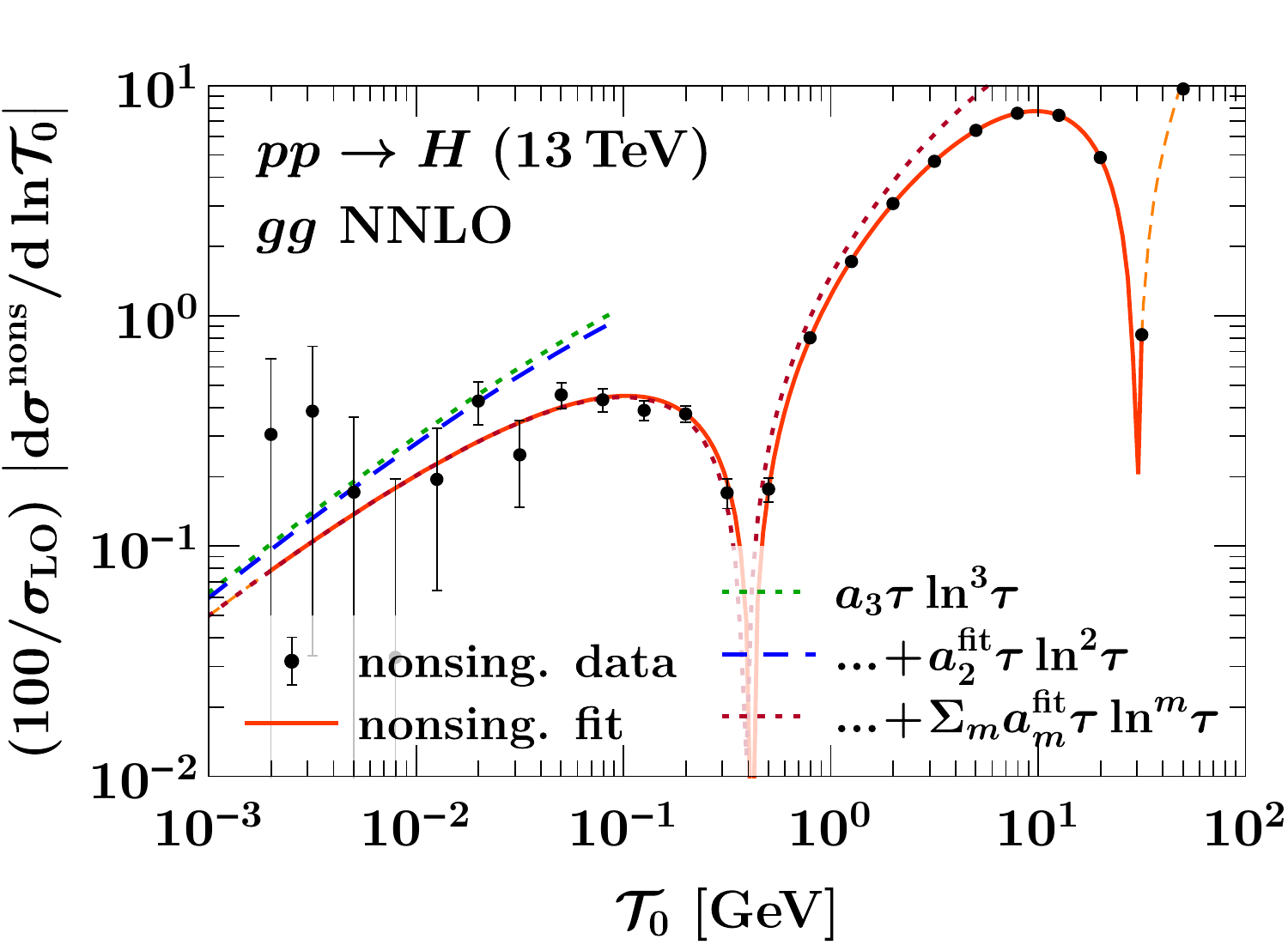}%
\\
\includegraphics[width=\columnwidth]{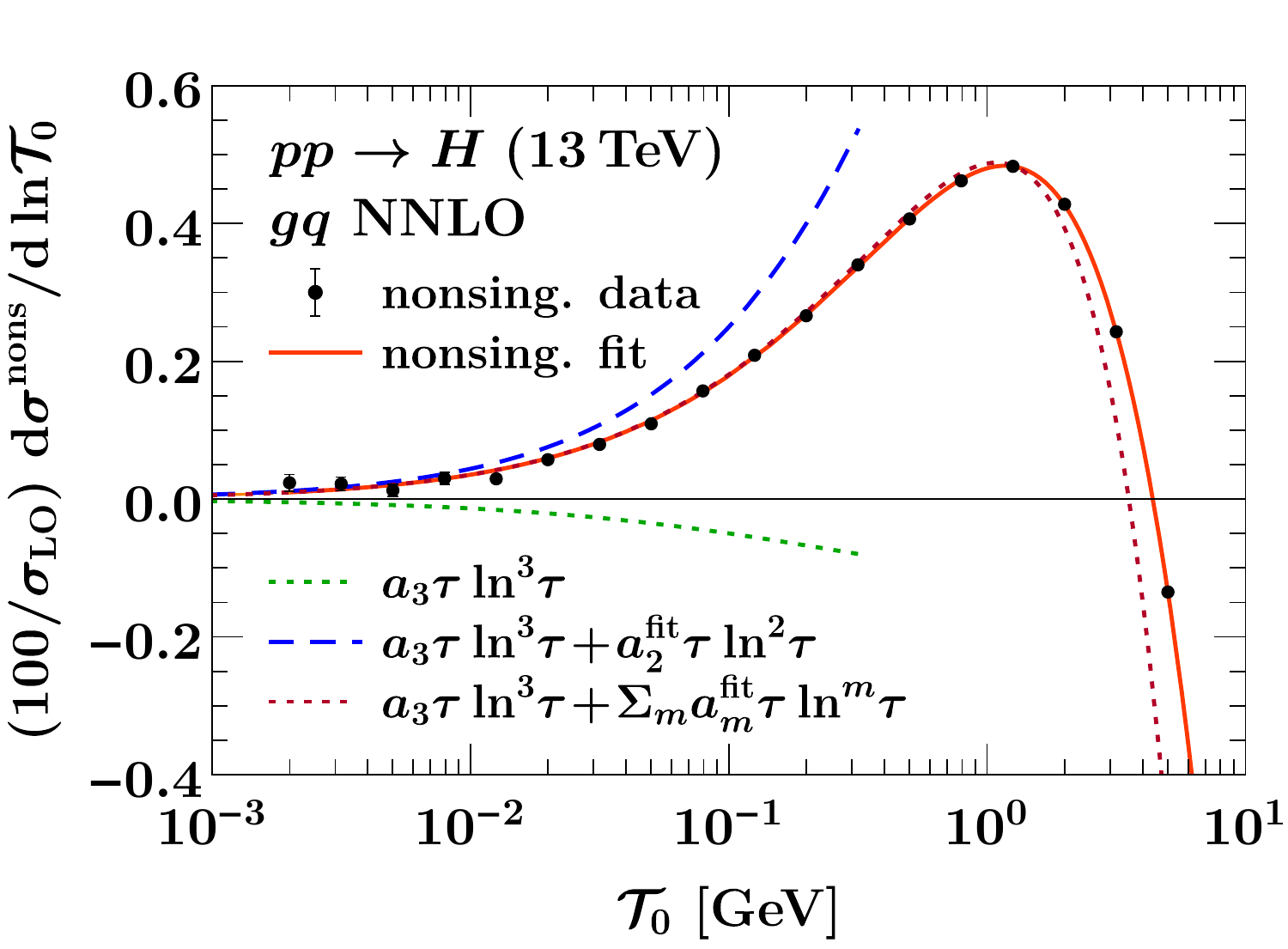}%
\hfill
\includegraphics[width=\columnwidth]{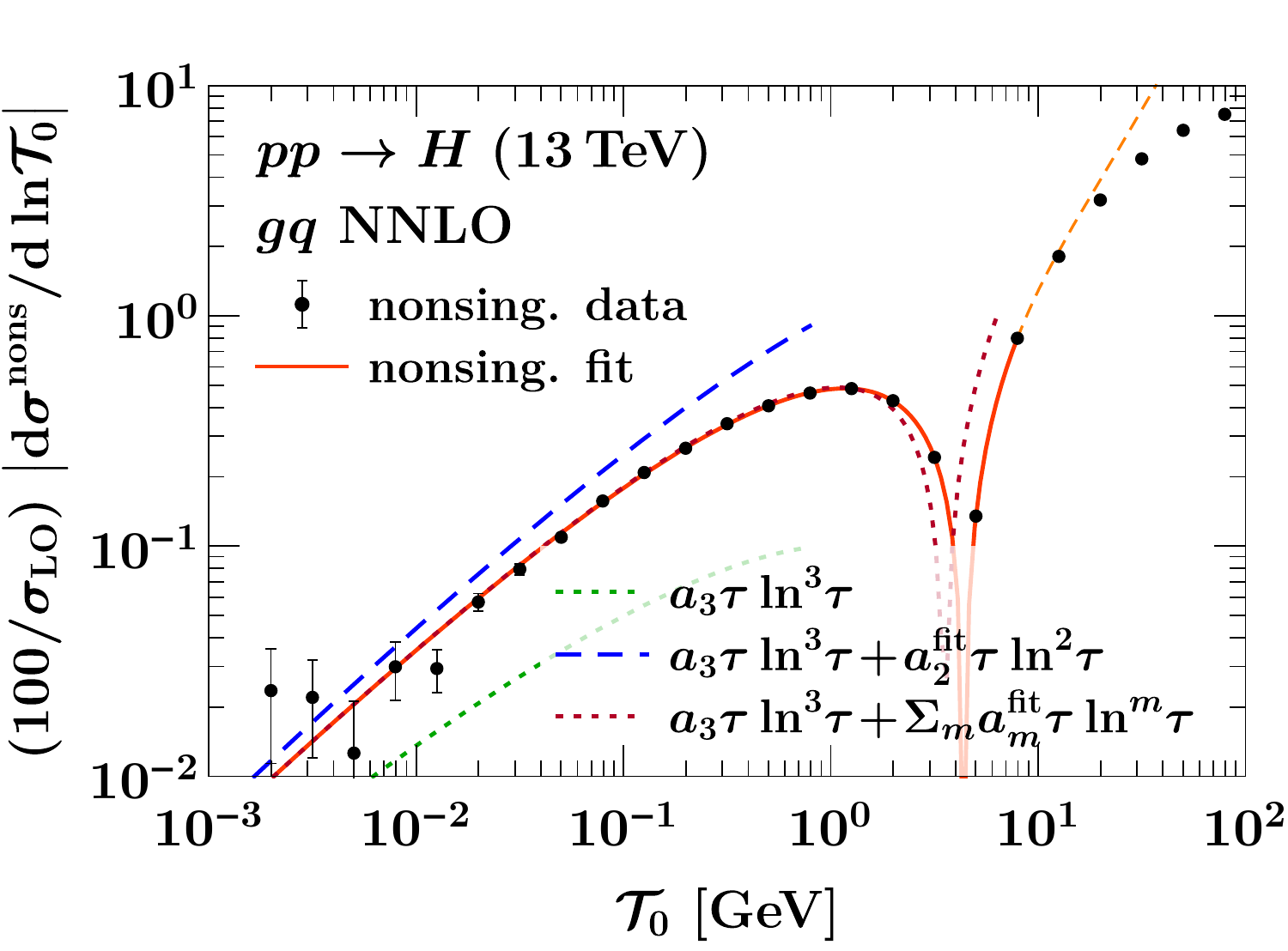}%
%%%
\caption{Fit to the $\ord{\alpha_s^2}$ nonsingular corrections for beam thrust in the $gg$ channel (top row) and the $gq$ channel (bottom row). The fit functions are defined in \eq{fitfun} and yield the solid red line (whose continuation by a light orange dashed line shows the extrapolation outside the fit region). The dashed blue line shows the result when only the leading coefficient from the red solid fit is retained, while the dotted red line shows the result when only the full set of logarithms at this power are retained. The plots in the right column are identical to those in the left column, but show the absolute value on a logarithmic scale.}
\label{fig:fitNNLO}
\end{figure*}

\begin{figure*}[t!]
\includegraphics[width=\columnwidth]{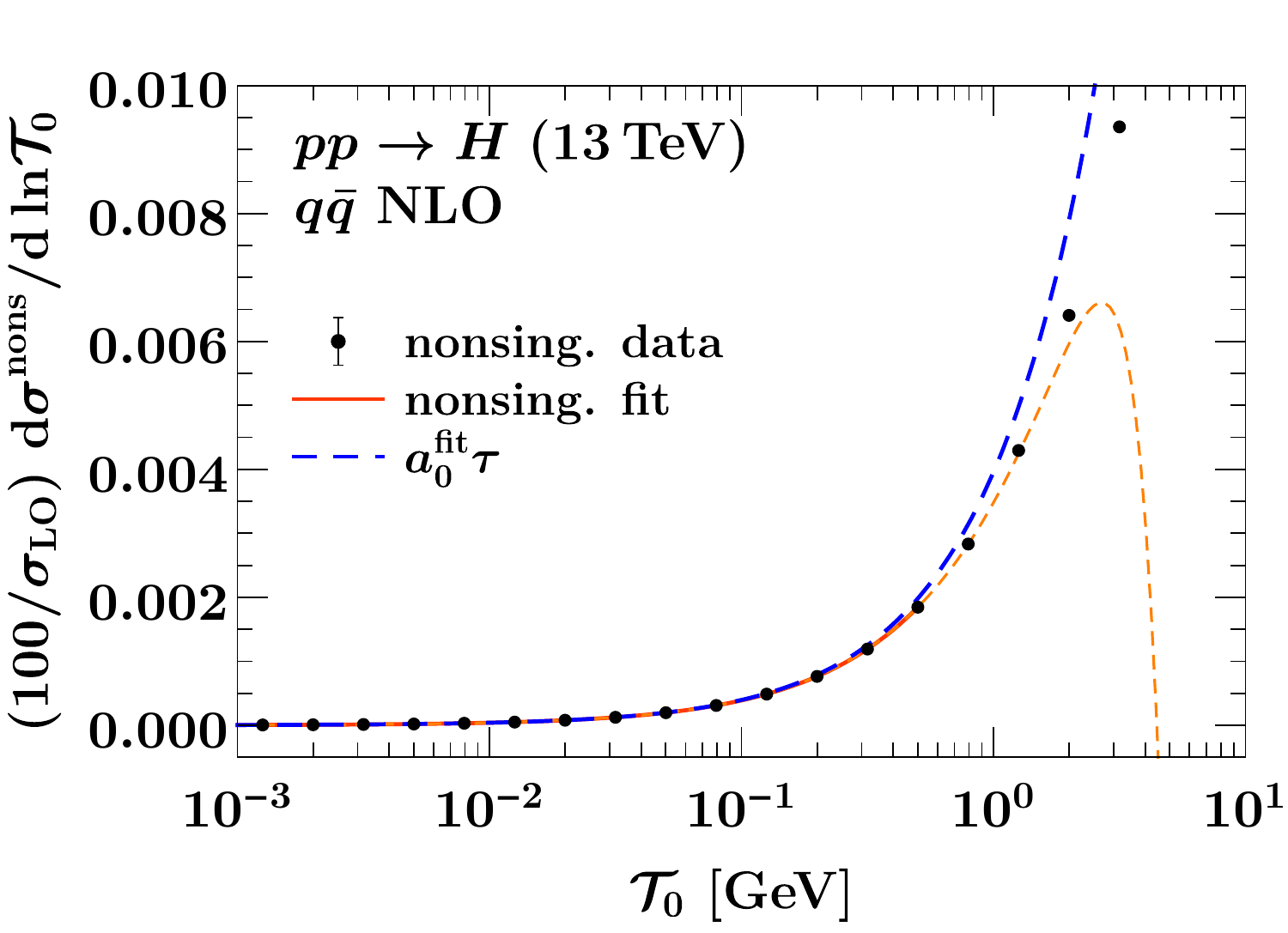}%
\hfill
\includegraphics[width=\columnwidth]{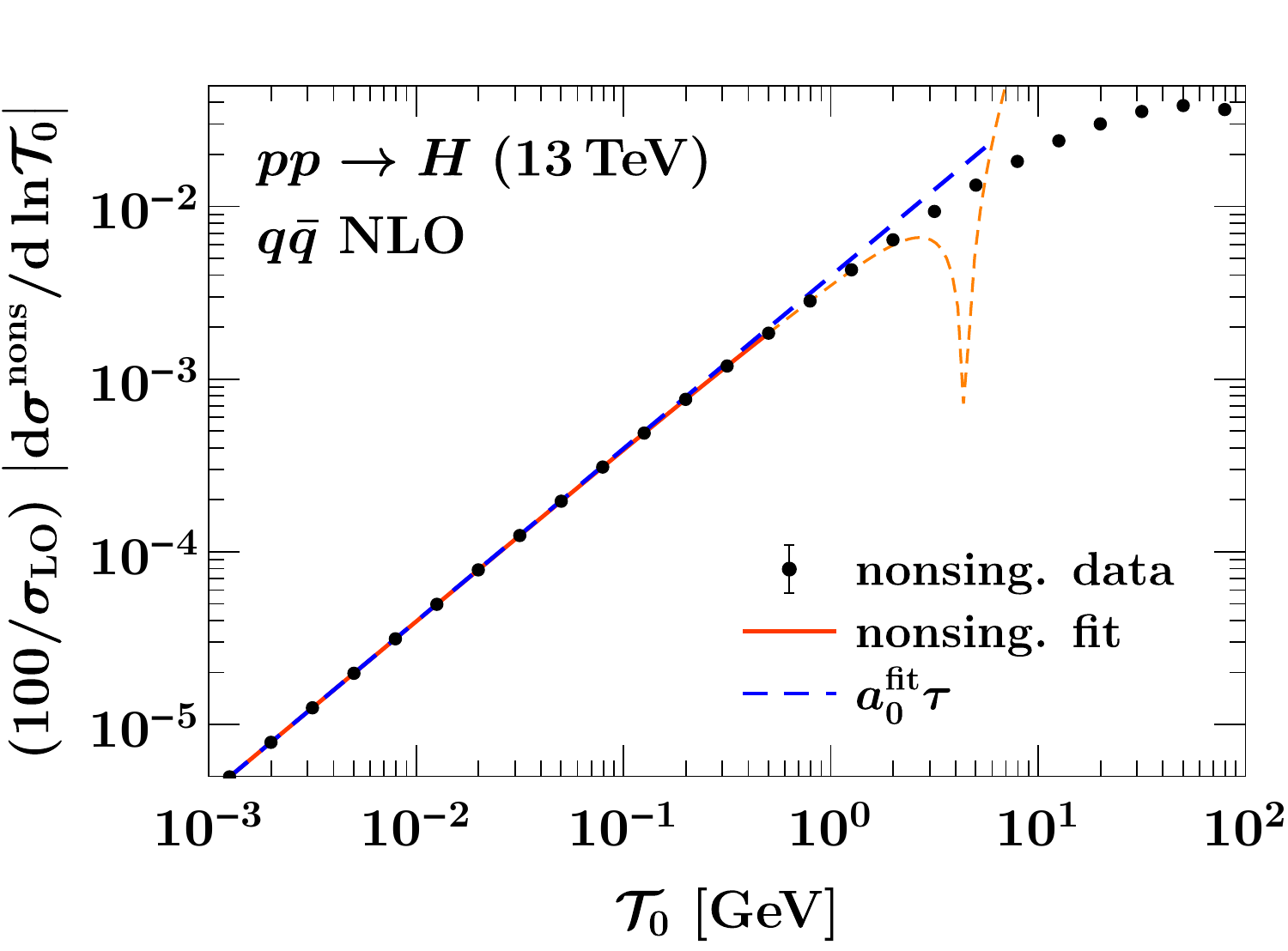}%
\\
\includegraphics[width=\columnwidth]{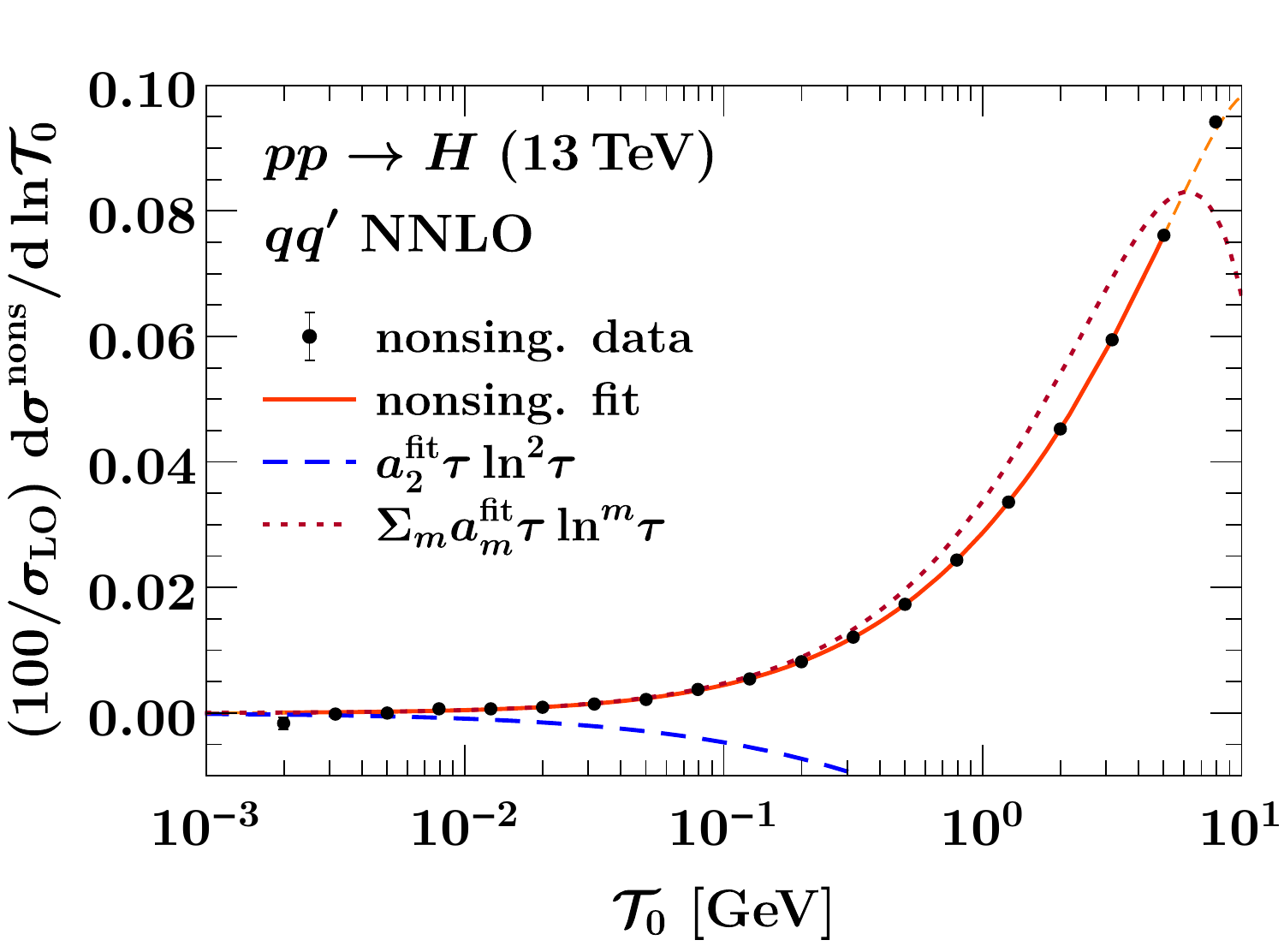}%
\hfill
\includegraphics[width=\columnwidth]{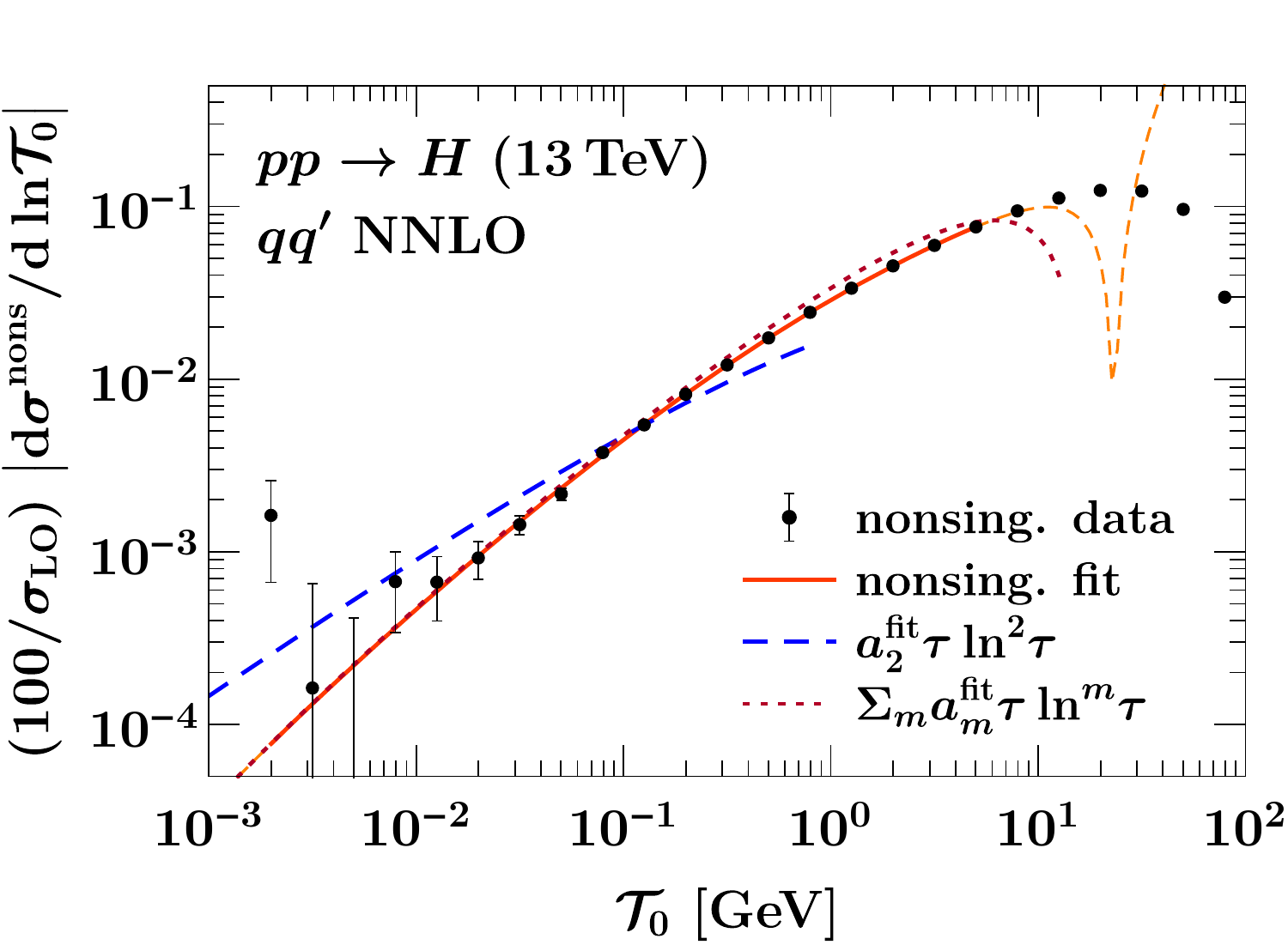}%
%%%
\caption{Fit to the nonsingular corrections for beam thrust in the $qq'$ channel. Results are shown at NLO (top row) and NNLO (bottom row). The fit functions are defined in \eq{fitfun} and yield the solid red line (whose continuation by a light orange dashed line shows the extrapolation outside the fit region). The dashed blue line shows the result when only the leading coefficient from the red solid fit is retained, while the dotted red line shows the result when only the full set of logarithms at this power are retained. The plots in the right column are identical to those in the left column, but show the absolute value on a logarithmic scale.}
\label{fig:fit_qq}
\end{figure*}

\begin{figure*}[t]
\includegraphics[width=\columnwidth]{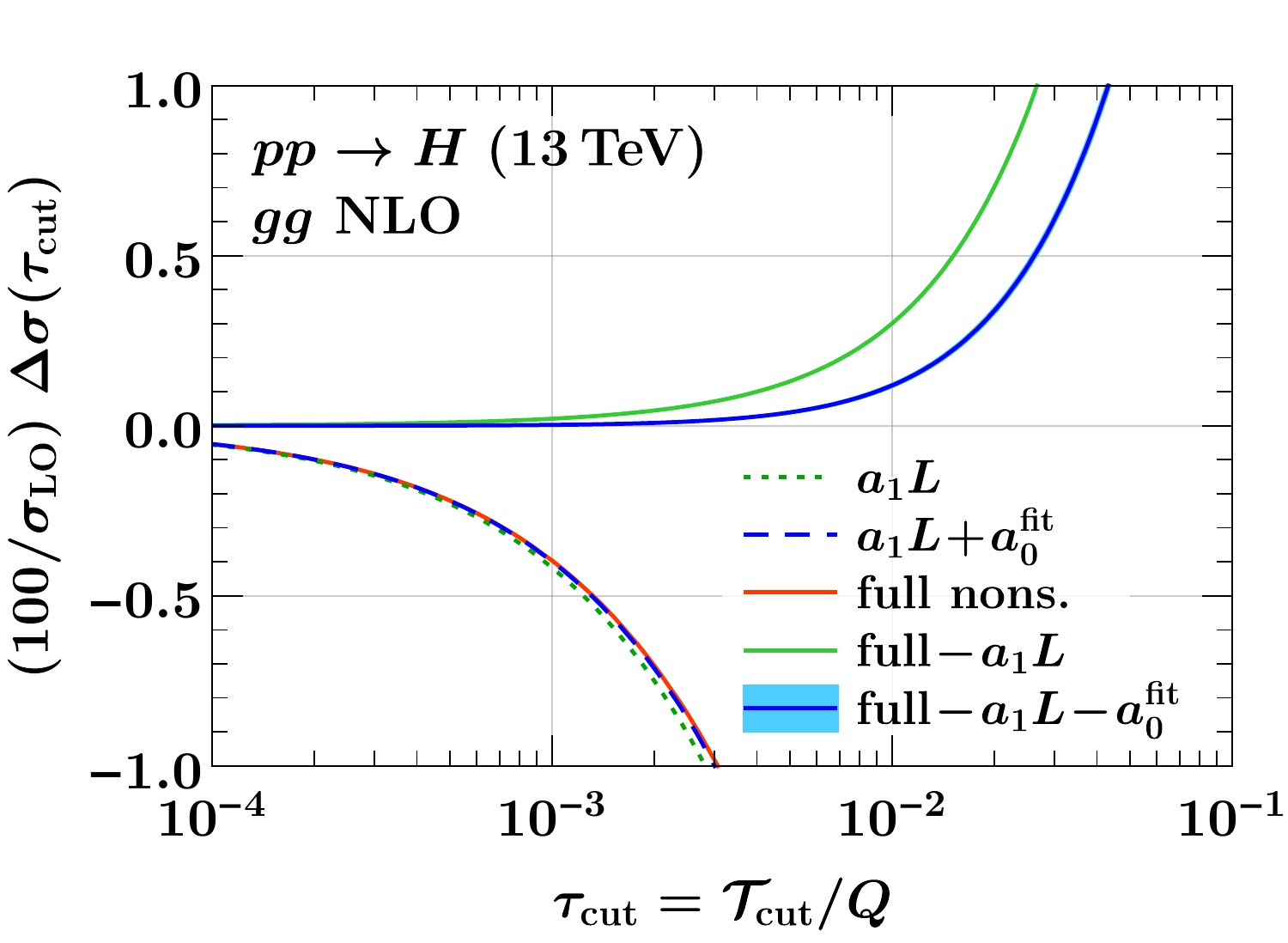}%
\hfill
\includegraphics[width=\columnwidth]{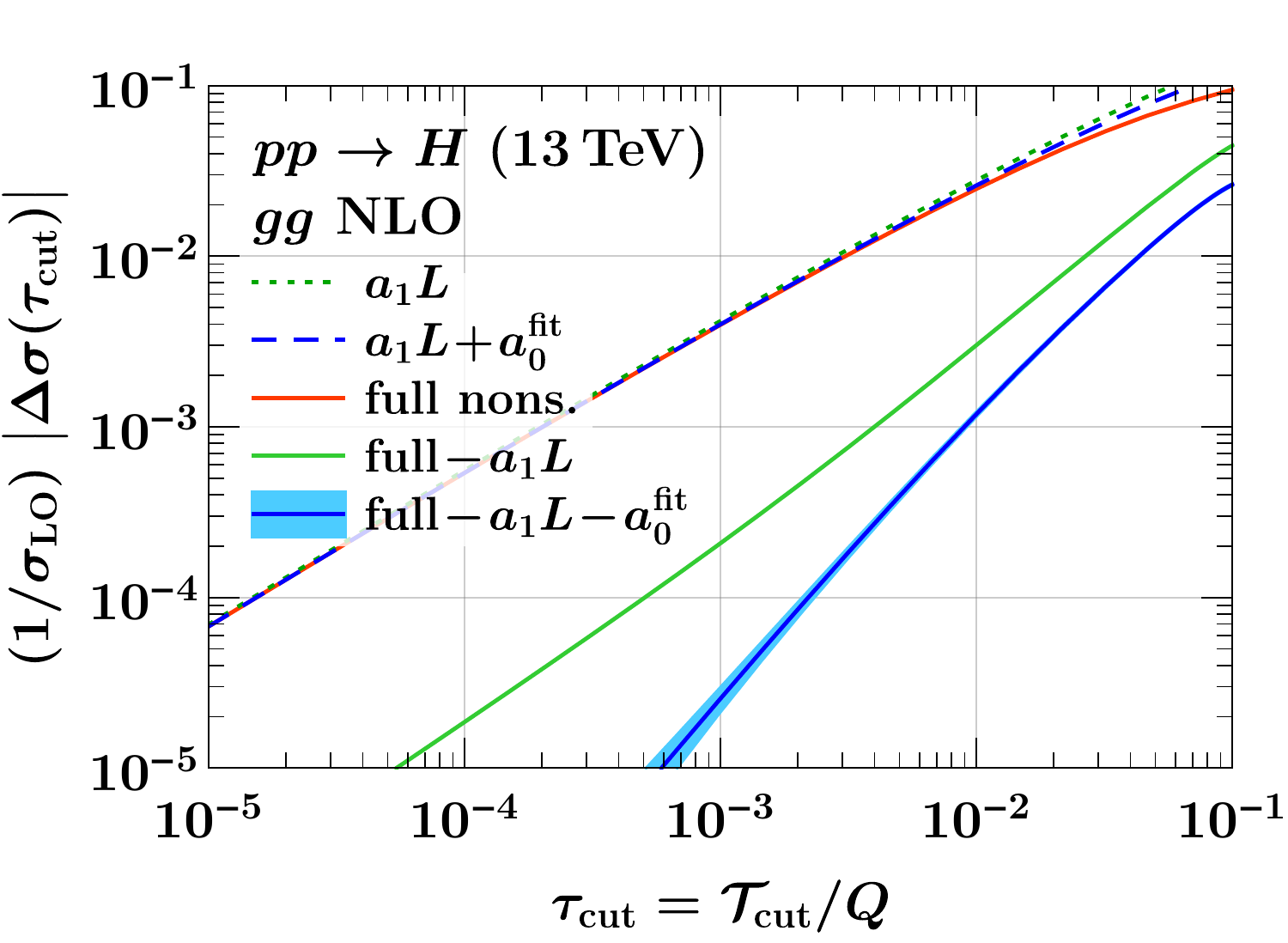}%
\\
\includegraphics[width=\columnwidth]{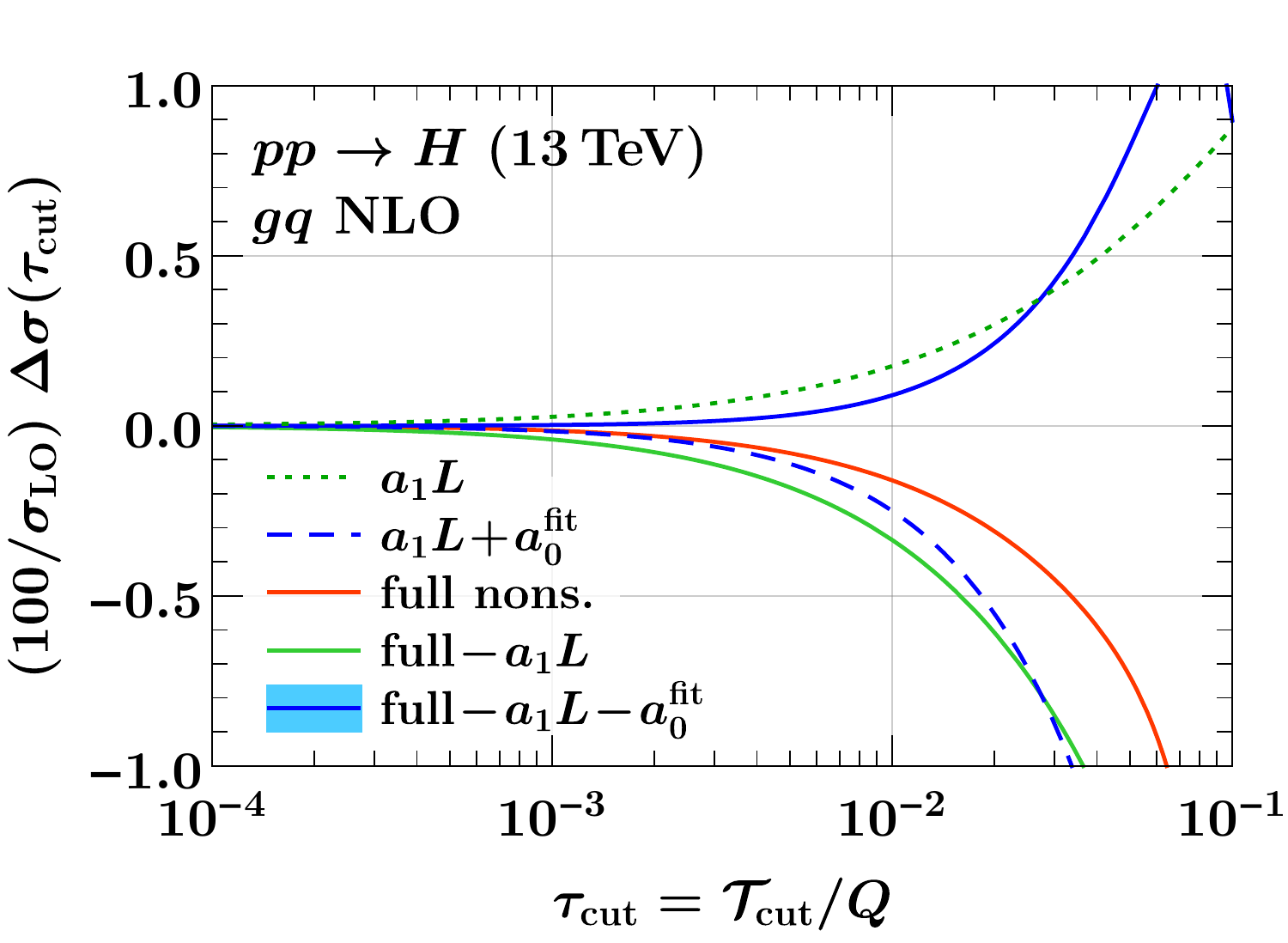}%
\hfill
\includegraphics[width=\columnwidth]{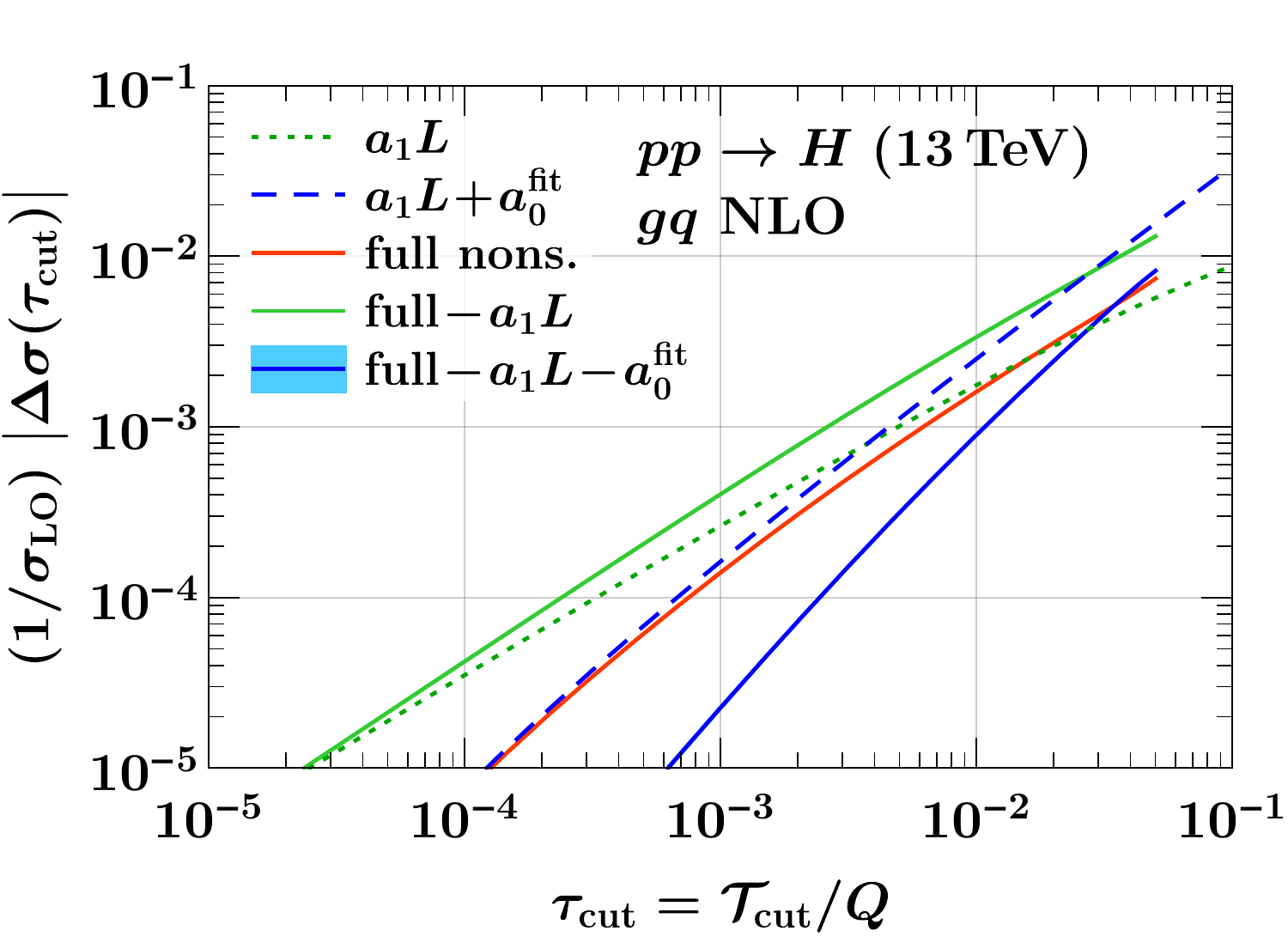}%
\\
\includegraphics[width=\columnwidth]{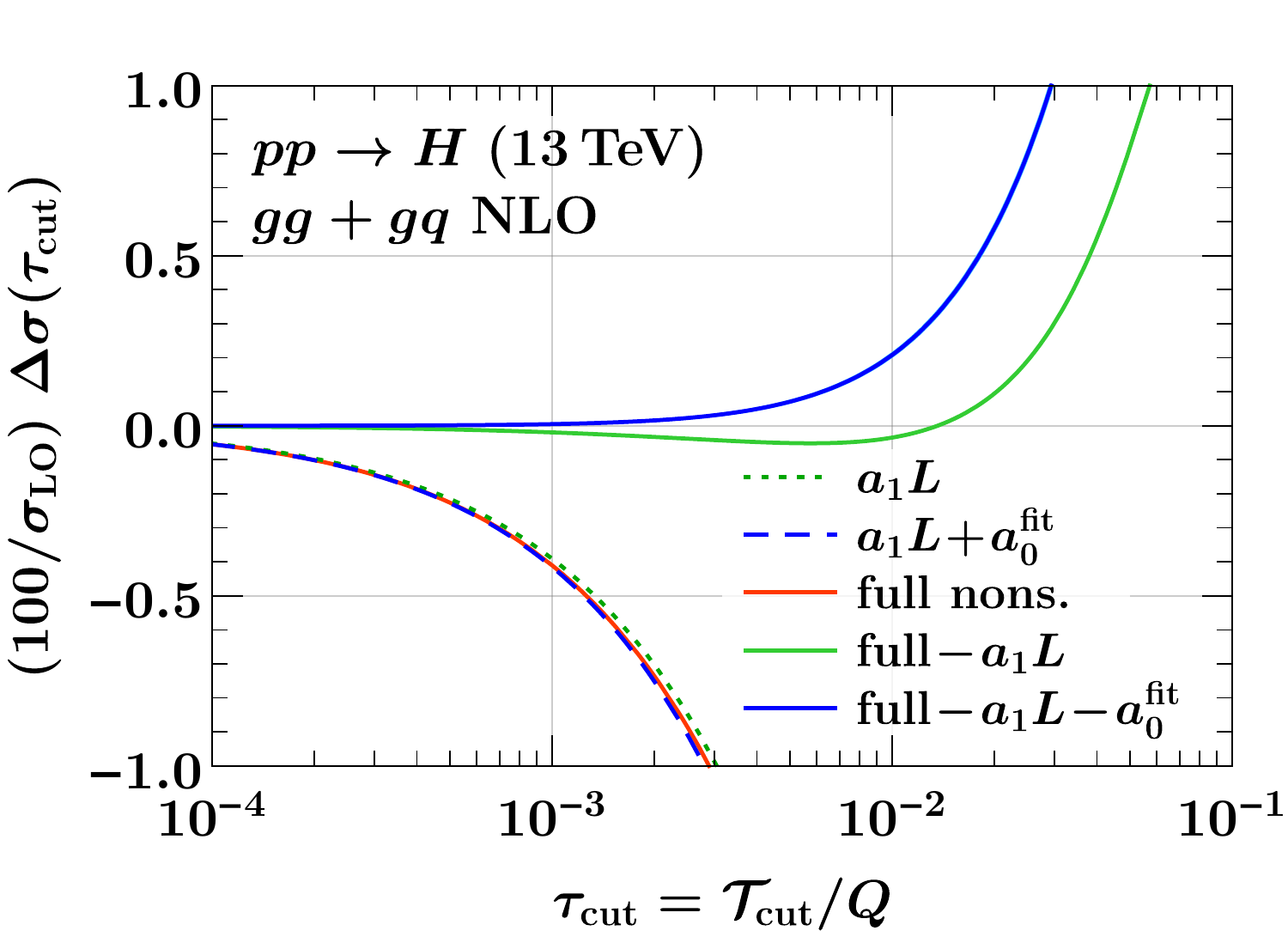}%
\hfill
\includegraphics[width=\columnwidth]{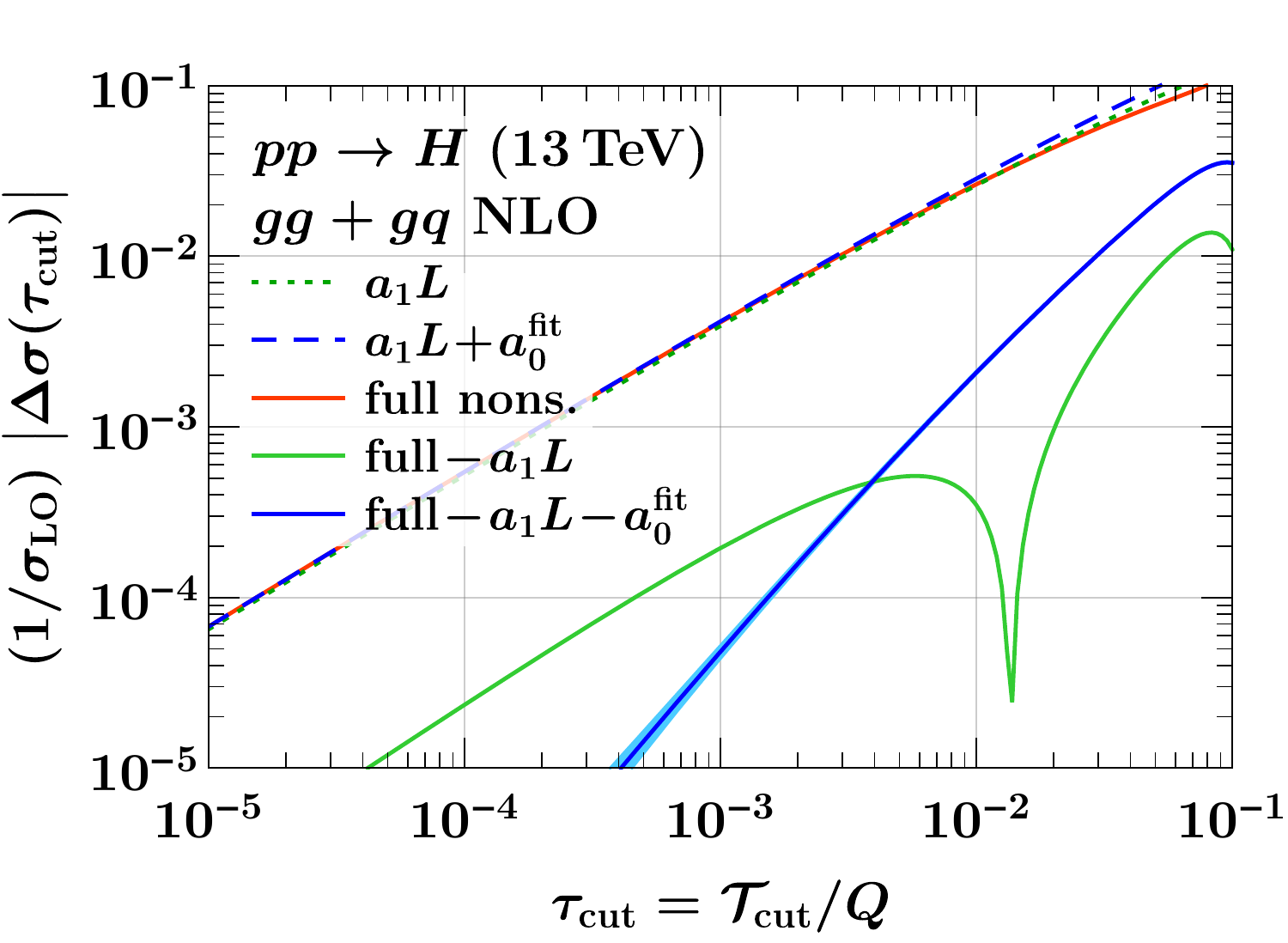}%
%%%
\caption{Power corrections $\Delta\sigma(\tau_\cut)$ for the $\ord{\alpha_s}$ contributions in the $gg$ channel (top row), the $gq$ channel (middle row), and the sum of both channels (bottom row). The plots in the right column are identical to those in the left column, but show the absolute value on a logarithmic scale.}
\label{fig:cumulantNLO}
\end{figure*}
\begin{figure*}[t]
\includegraphics[width=\columnwidth]{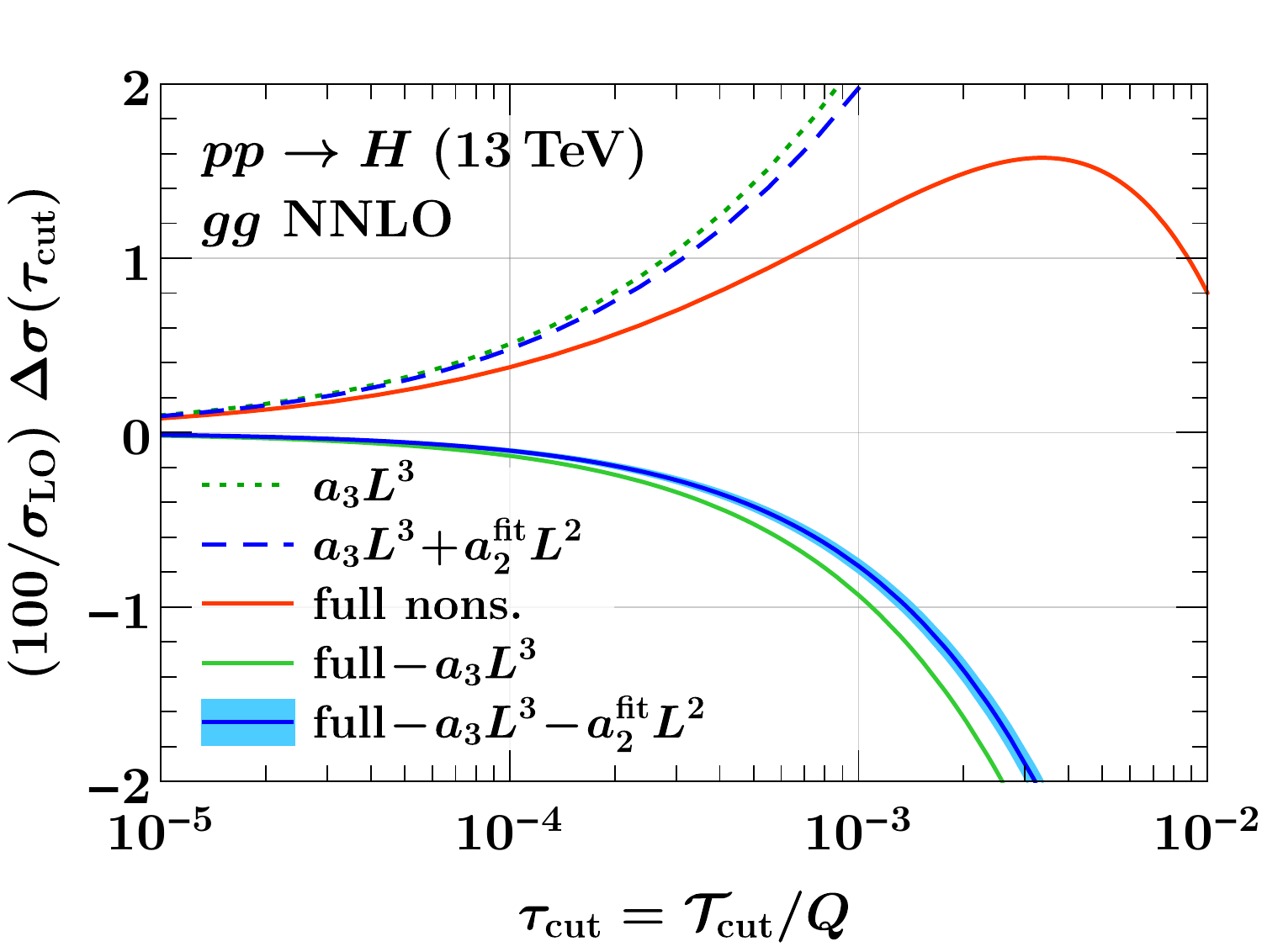}%
\hfill
\includegraphics[width=\columnwidth]{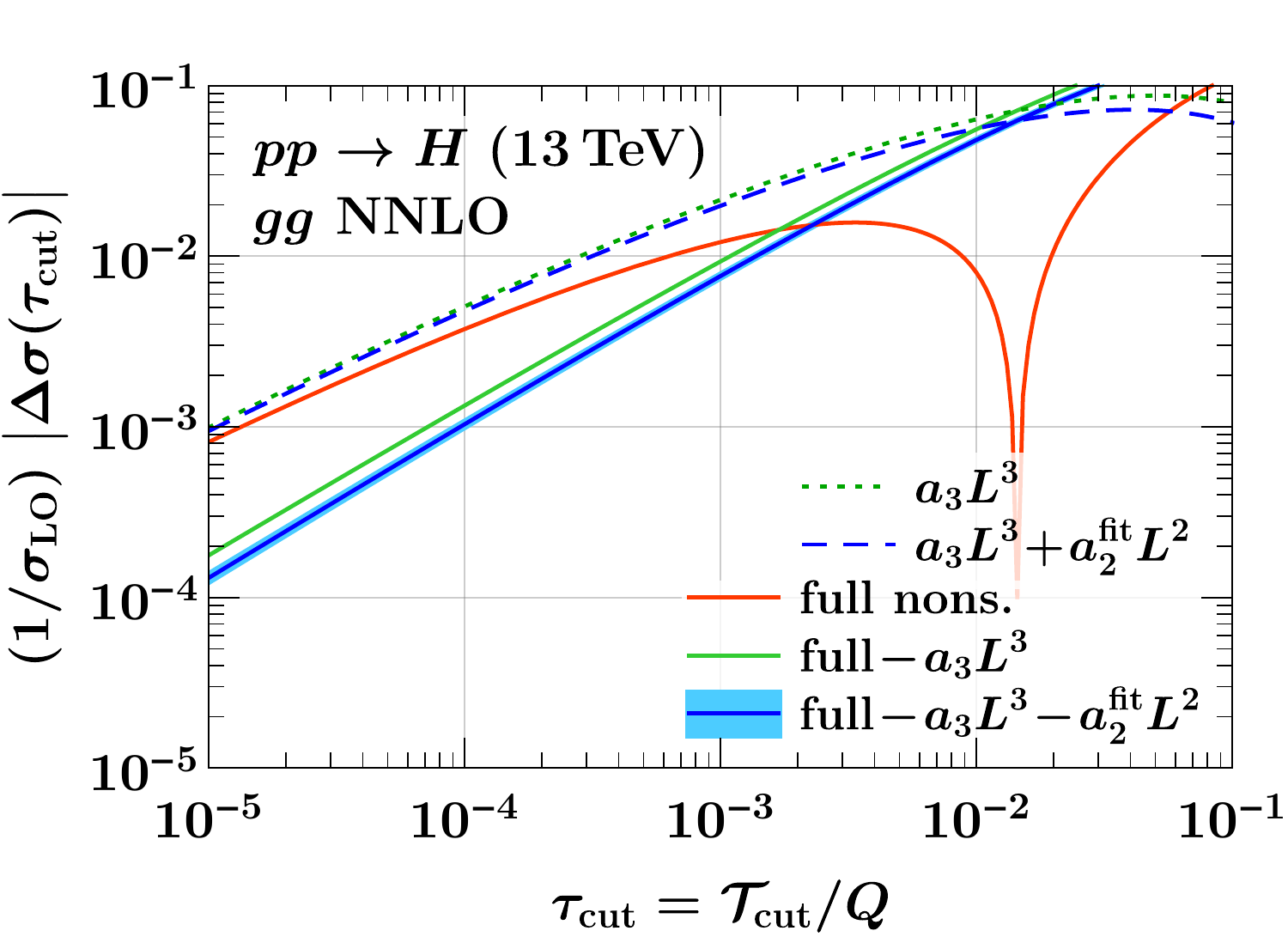}%
\\
\includegraphics[width=\columnwidth]{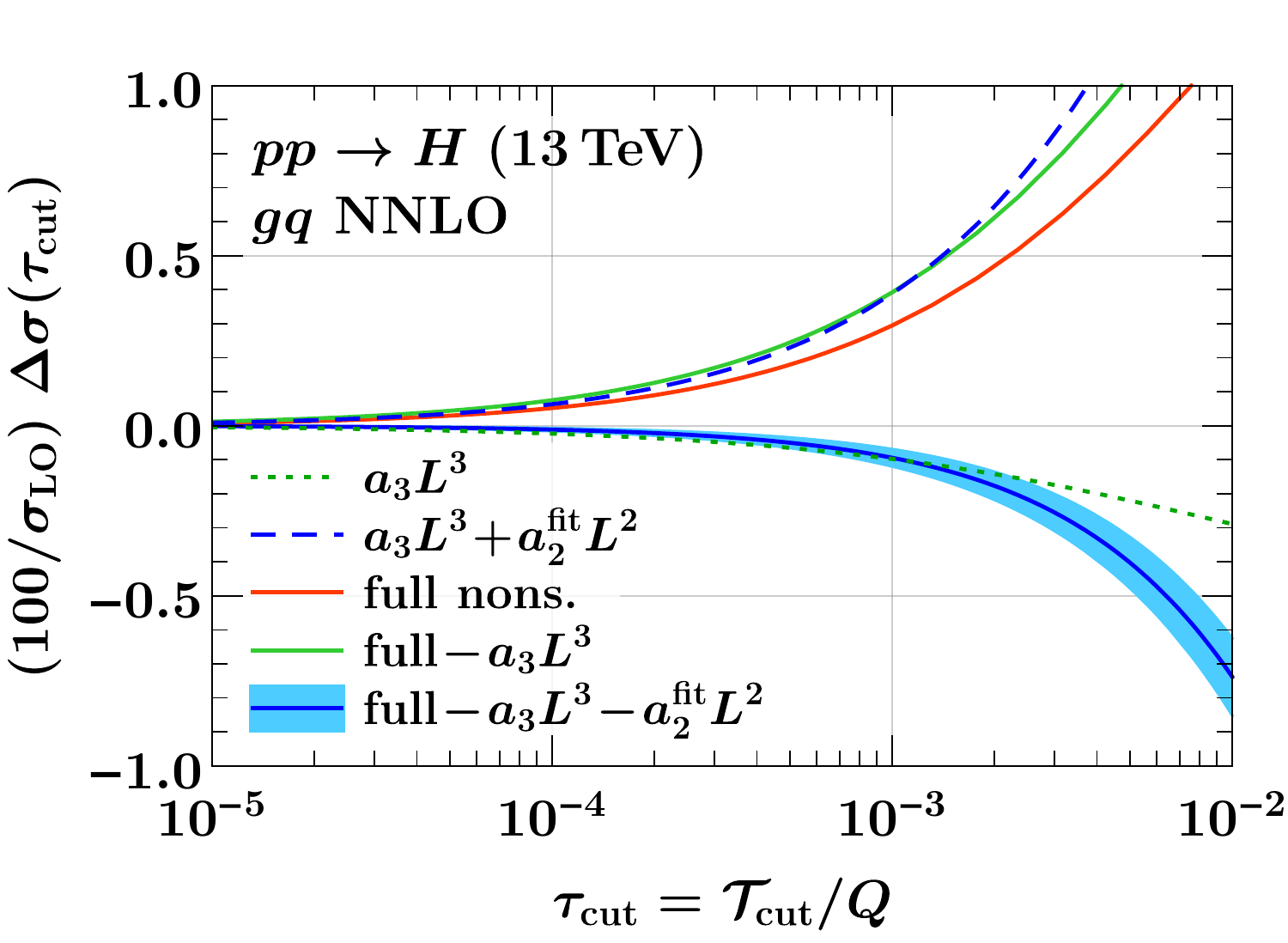}%
\hfill
\includegraphics[width=\columnwidth]{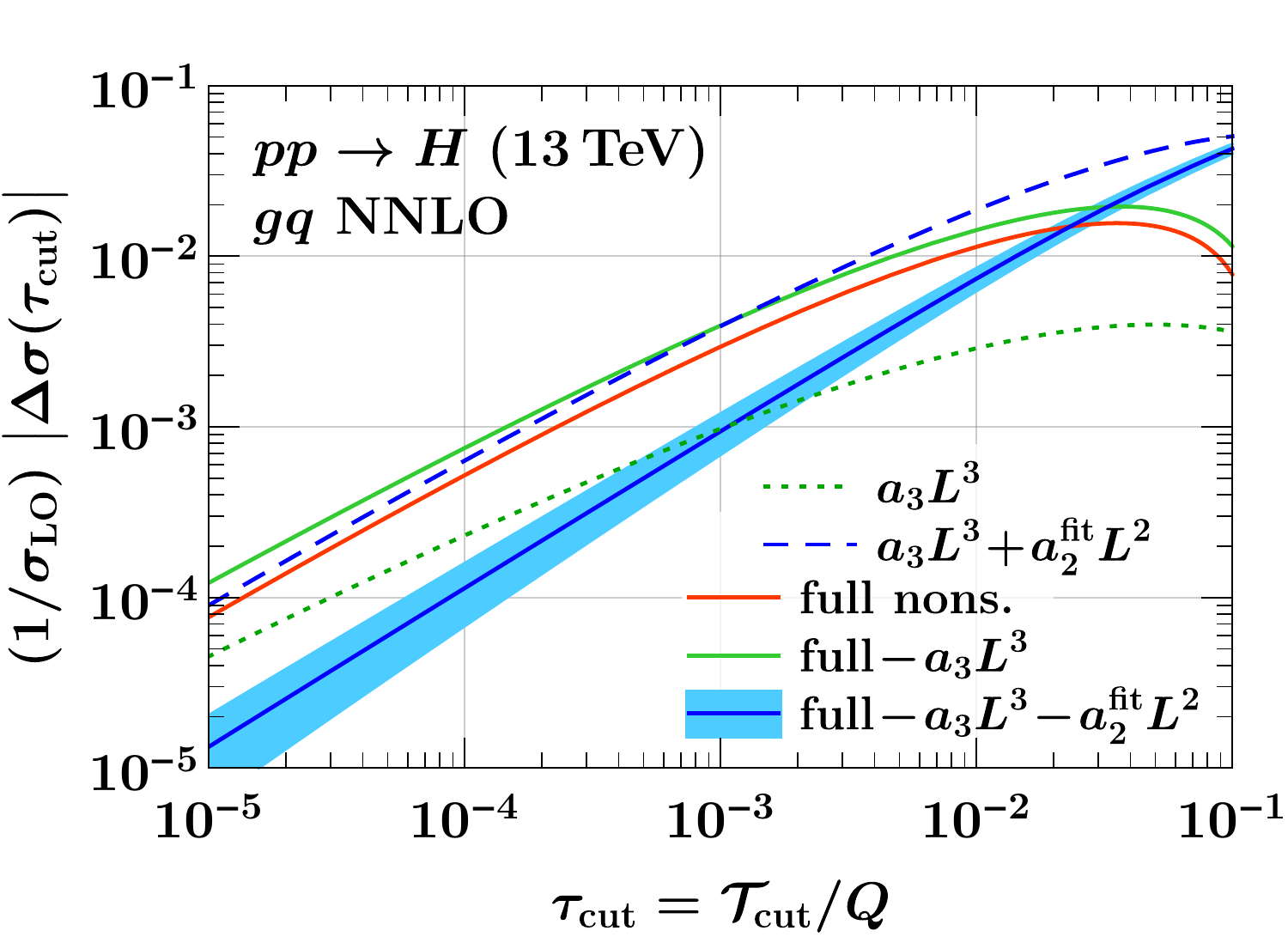}%
\\
\includegraphics[width=\columnwidth]{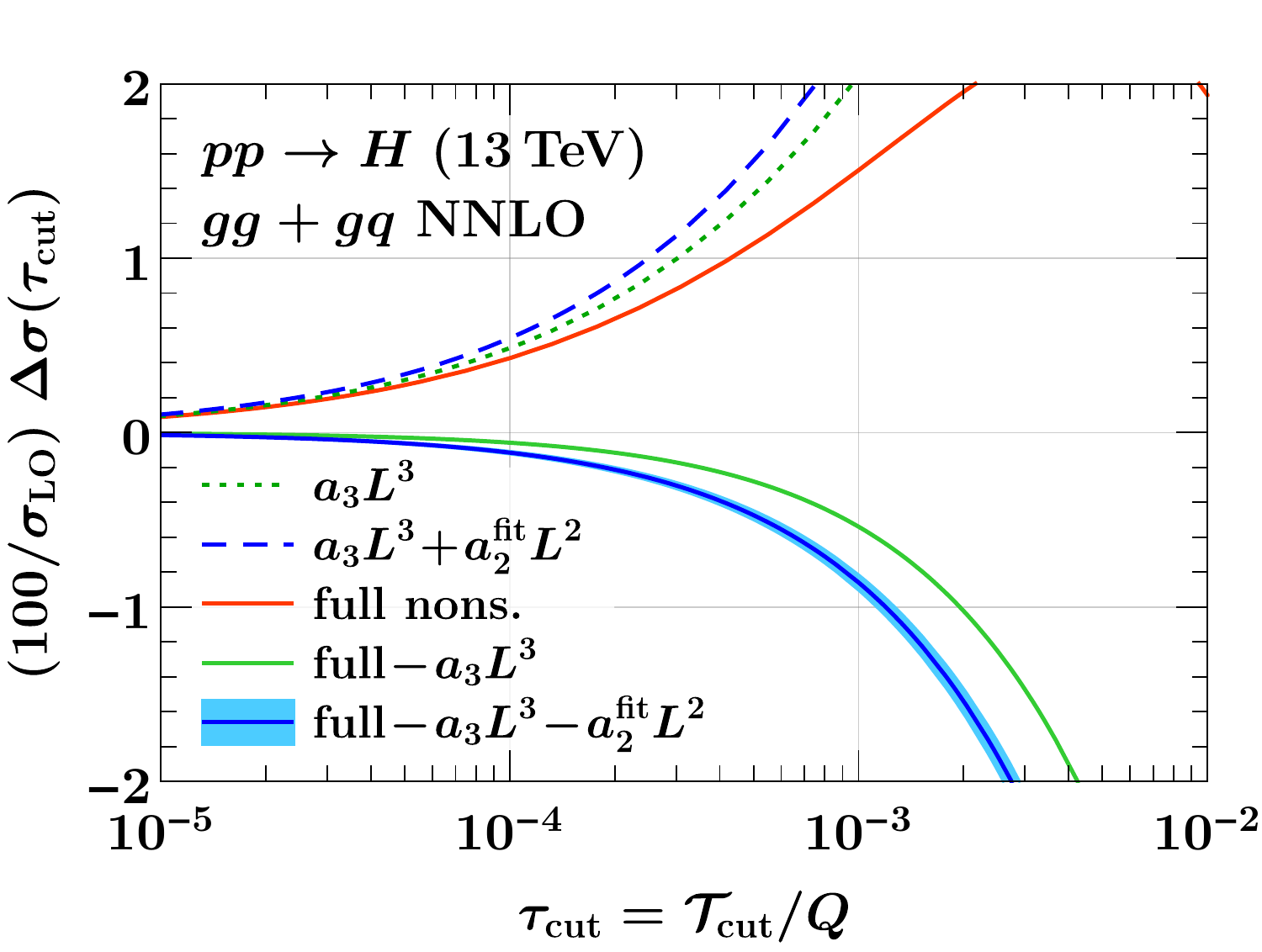}%
\hfill
\includegraphics[width=\columnwidth]{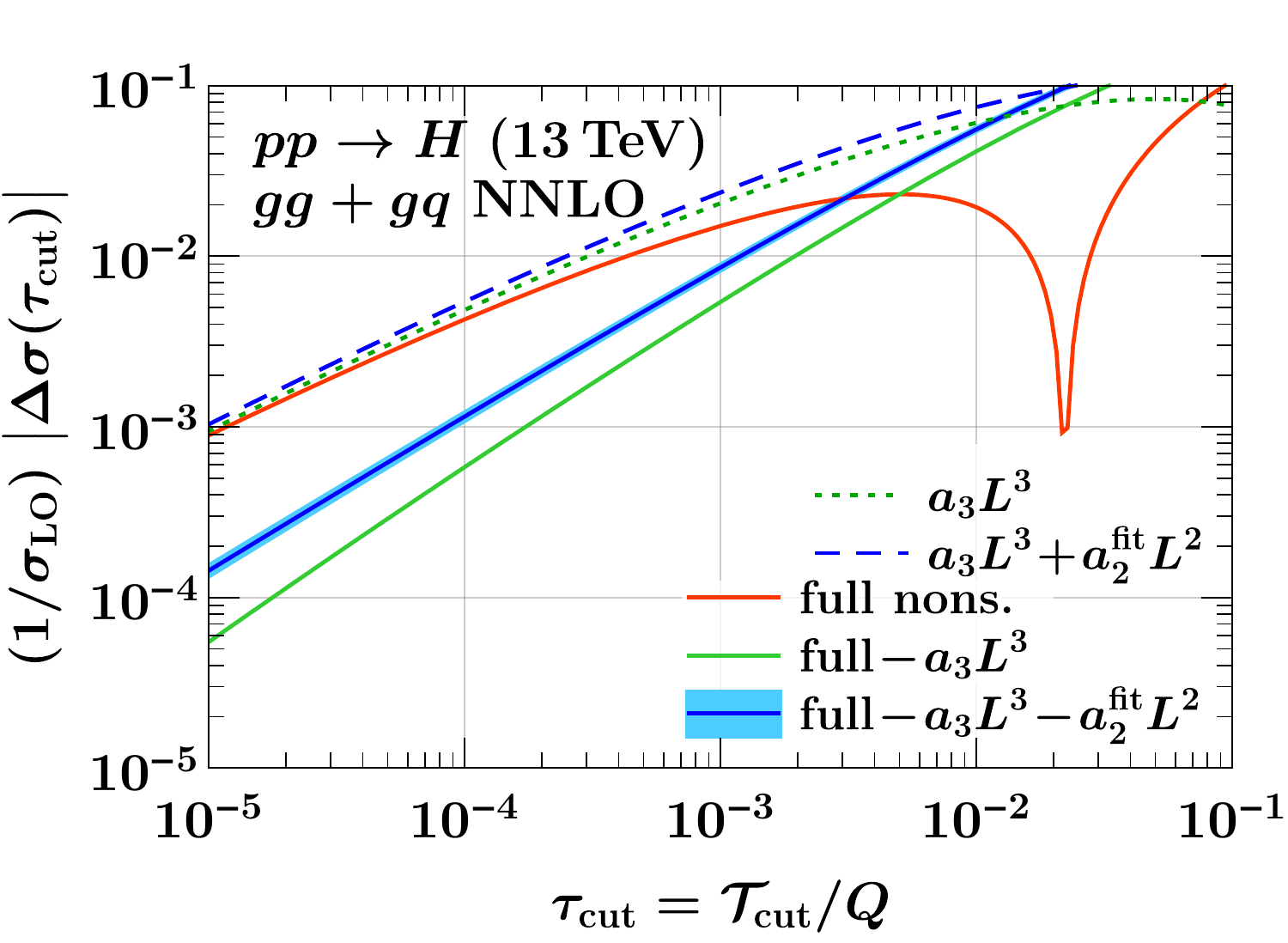}%
%%%
\caption{Power corrections $\Delta\sigma(\tau_\cut)$ for the $\ord{\alpha_s^2}$ contributions in the $gg$ channel (top row), the $gq$ channel (middle row), and the sum of both channels (bottom row).
The plots in the right column are identical to those in the left column, but show the absolute value on a logarithmic scale.}
\label{fig:cumulantNNLO}
\end{figure*}

In this section, we compare our leptonic $\tau$ results with numerical fixed-order results.
This first serves as a numerical cross check on our
calculated coefficients for the leading logarithm at subleading power. Then, using our exact
results for the LL coefficients allows us to use the numerical results to extract the NLL corrections
of $\alpha_s \tau$ and $\ord{\alpha_s^2\tau \ln^2\tau}$, and to study the
relative importance of the LL and NLL contributions.

We consider the process $pp\to H$
at $E_{\rm cm} = 13\TeV$ for an on-shell, stable Higgs boson with $m_H = 125\GeV$.
We always use the MMHT2014 NNLO PDFs~\cite{Harland-Lang:2014zoa},
fixed scales $\mu_r = \mu_f = m_H$, and $\alpha_s(m_H) = 0.1126428$.
The leading-order cross section following from \eq{sigmapartLO} is given by
%%%
\begin{equation}
\sigma_\LO
= \sigma_0\int\!\df Y f_g\bigl(m_H e^Y\bigr) f_g\bigl(m_H e^{-Y}\bigr)
\,.\end{equation}
%%%
We normalize all our results to $\sigma_\LO$, as it provides the most natural common
reference value to discuss the size of the power corrections in the different  channels.
This also removes most of the dependence on the explicit process mediated by the $ggH$ operator.
The only leftover dependence is related to the PDFs and comes from the effective $x$-range in the PDFs probed by the rapidity integration, which is determined by the value of $m_H$. The PDF dependence primarily determines the relative
size of the $gg$, $gq$, and $q\bar q$ channels, but only to a small extent the size of the corrections within a given channel.

We obtain the full $\Tau_0$ spectrum at $\ord{\alpha_s}$ and $\ord{\alpha_s^2}$ using the $H+1$-jet (N)LO
calculation from MCFM8~\cite{Campbell:1999ah, Campbell:2010ff, Campbell:2015qma, Boughezal:2016wmq}.
The known leading-power terms in the $\Tau_0$ spectrum~\cite{Stewart:2009yx, Stewart:2010qs, Berger:2010xi} are then subtracted to obtain the complete nonsingular (all subleading-power) contributions,
%%%
\begin{equation} \label{eq:dsigmanons}
\frac{1}{\sigma_\LO}\frac{\df\sigma^\nons}{\df\ln\Tau_0}
= \frac{1}{\sigma_\LO}\frac{\df\sigma}{\df\ln\Tau_0} - \frac{1}{\sigma_\LO}\frac{\df\sigma^{(0)}}{\df\ln\Tau_0}
\,.\end{equation}
%%%
We do this separately for the $\alpha_s$ (NLO) and pure $\alpha_s^2$ (NNLO) contributions and separately for the $gg$, $gq$, and $q\bar q$ (or $qq'$) channels. The $gq$ channel includes the sum of the $gq$ and $qg$ contributions with $q$ summed over all quarks and antiquarks. The $q\bar q$ (or $qq'$) channel includes the sum over all allowed flavor combinations.

Our numerical analysis follows the same fit strategy as for the case of Drell-Yan in Ref.~\cite{Moult:2016fqy}.
We fit the nonsingular NLO and NNLO data in each partonic channel using the functional forms
%%%
\begin{align} \label{eq:fitfun}
F_\mathrm{NLO}(\tau)
&= \frac{\df}{\df\ln\tau}\Bigl\{ \tau \bigl[(a_1 + b_1 \tau + c_1 \tau^2) \ln\tau
\nn\\ & \quad
+ a_0 + b_0\tau + c_0\tau^2 \bigr] \Bigr\}
\,,\nn\\
F_\mathrm{NNLO}(\tau)
&= \frac{\df}{\df\ln\tau}\Bigl\{\tau \bigl[ (a_3 + b_3 \tau) \ln^3\tau + (a_2 + b_2\tau) \ln^2\tau
\nn\\ & \quad
+ a_1 \ln\tau + a_0 \bigr] \Bigr\}
\,,\end{align}
%%%
with $\tau \equiv \Tau_0/m_H$.
Since the different powers of $\ln\tau$ have very similar shapes, the fitted coefficients
at the same order in $\tau$ are typically highly correlated and care has to be taken to
ensure reliable and unbiased fit results. An important consideration in this regard is
the choice of fit range in $\Tau_0$ and the number of fit coefficients.
We refer to Ref.~\cite{Moult:2016fqy} for a detailed discussion of these issues.

\begin{table}[h]
\centering
\begin{tabular}{cc|cc}
\hline\hline
\multicolumn{2}{c|}{order and channel} & fitted & calculated
\\ \hline
NLO $gg$ & $a_1$ & $+0.60936 \pm 0.00600$ & $+0.60400$
\\
NLO $gq$ & $a_1$ & $-0.03733 \pm 0.00066$ & $-0.03807$
\\ 
NLO $q\bar q$ & $a_1$ & $(1.53 \pm 1.62)\times 10^{-7} $ & $0$
\\ 
NLO $q\bar q$ & $10^3 a_0$ & $+4.90060 \pm 0.00013$ & $+4.90048$
\\
\hline
NNLO $gg$ & $a_3$ & $-0.05785 \pm 0.00713$ & $-0.06497$
\\
NNLO $gq$ & $a_3$ & $+0.00998 \pm 0.00509$ & $+0.00296$
\\
NNLO $qq'$ & $a_3$ & $+0.00021 \pm 0.00019$ & $0$
\\
\hline\hline
\end{tabular}
\caption{Comparison of fitted and calculated values for the LL coefficients. For the $q\bar q$ channel at NLO we include both the LL coefficient, which we confirm to be zero, and the nonvanishing NLL coefficient, which we also computed. In all cases we find
excellent agreement between fitted and calculated results.}
\label{tab:LLresults}
\end{table}

\begin{table}[h]
\centering
\begin{tabular}{cc|c}
\hline\hline
\multicolumn{2}{c|}{order and channel} & fitted
\\ \hline
NLO $gg$ & $a_0$ & $+0.18241 \pm 0.00425$
\\
NLO $gq$ & $a_0$ & $-0.42552 \pm 0.00032$
\\
 \hline
NNLO $gg$ & $a_2$ & $-0.03491 \pm 0.00758$
\\
NNLO $gq$ & $a_2$ & $+0.10193 \pm 0.00536$
\\
NNLO $qq'$ & $a_2$ & $-0.00159\pm 0.00037$
\\ \hline\hline
\end{tabular}
\caption{Fit results for the NLL coefficients using the calculated LL coefficients in table~\ref{tab:LLresults} as input. At NLO we have not included the $q\bar q$ channel, since this channel starts at subleading logarithmic order, and the $a_0$ coefficient is analytically calculated and given in table~\ref{tab:LLresults}. }
\label{tab:NLLresults}
\end{table}

As a check of our calculation we first extract the LL coefficients from the fit, which are $a_1$ at NLO and $a_3$ at NNLO. The results are given in table~\ref{tab:LLresults} along with the values predicted from our calculation. In all cases we find excellent agreement.
We then perform the fit with the LL coefficients fixed to their calculated values, which allows for a precise extraction of the NLL coefficients $a_0$ at NLO and $a_2$ at NNLO. The results are shown in table~\ref{tab:NLLresults}, and the corresponding fits are illustrated in Figs.~\ref{fig:fitNLO}, \ref{fig:fitNNLO}, and \ref{fig:fit_qq}.

As discussed above, for the channels involving only quarks, we have only $q\bar q$ at NLO, but $q q'$ at NNLO, where $q'$ is an arbitrarily flavored quark or antiquark, and includes the particular case of $q\bar q$. For the $q\bar q$ and $qq'$ channels, from our calculation we find $a_1=a_3=0$. At NLO, we are able to verify to high accuracy that $a_1=0$ from the fit, and extract $a_0$ to compare with our calculated value. At NNLO, the result for $a_3$ in this channel is consistent with zero.

If we approximate the $\xi_a$ and $\xi_b$ dependence of the NLL coefficients in the partonic cross section by the corresponding dependence at LL, we can translate the fitted values for $a_0$ and $a_2$ into the approximate results
%%%
\begin{align}
C_{gg, 0}^{(2,1)}(\xi_a, \xi_b)
&\approx (31.2\pm 0.2) \biggl(\delta_a \delta_b+\frac{\delta_a' \delta_b}{2} + \frac{\delta_a \delta_b'}{2}\biggr)
\,, \nn \\
C_{gq, 0}^{(2,1)}(\xi_a, \xi_b)
&\approx -32.5\,\delta_a \delta_b
\,, \nn \\
C_{gg, 2}^{(2,2)}(\xi_a, \xi_b)
&\approx (-1019 \pm 34) \biggl(\delta_a \delta_b+\frac{\delta_a' \delta_b}{2} +\frac{\delta_a \delta_b'}{2}\biggr)
\,, \nn \\
C_{gq, 2}^{(2,2)}(\xi_a, \xi_b)
&\approx (866 \pm 42) \delta_a \delta_b
\,,\end{align}
%%%
where the uncertainties arise from the fit uncertainties of the respective subleading coefficient. We do not give a result for the $qq'$ channel, since it is phenomenologically irrelevant, and more care must be taken since it is a sum over all possible quark channels and an approximate result can be easily misinterpreted. Nevertheless,  it can be obtained from our above results by carefully performing the sum over quark flavors.

At both NLO, \fig{fitNLO}, and NNLO, \fig{fitNNLO}, we see that there is a significant difference between the structure of the power corrections in the $gg$ and $gq$ partonic channels. In the $gg$ channel at each order, the NLL coefficients have the same sign and a comparable magnitude to the LL coefficients. Hence, the LL power correction provides a good first approximation to the total nonsingular correction. On the other hand, for the $gq$ channel, the LL and NLL coefficients have opposite signs, and the LL coefficients are very small compared to both their $gq$ NLL coefficients and also the $gg$ LL coefficients, while the NLL coefficients for $gq$ are of similar size as for $gg$. The smallness of the LL coefficient for $gq$ compared to $gg$ is due to its much smaller color factor. Our numerical results thus imply that the same color suppression will not be at work any longer at NLL.
This different behavior for the different channels motivates that it would be interesting to calculate the NLL coefficients analytically. This also means that the LL terms are by themselves a poor approximation to the full nonsingular result in the $gq$ channel. Indeed, in \figs{fitNLO}{fitNNLO}, the agreement between the LL result and the full non-singular contribution for the $gq$ channel (lower right panel) is poor in the range shown. Due to the fact that the NLL term is much larger, there will be a flip in the sign of the result at values of $\Tau_0$ lower than we are able to numerically probe. This behavior is more clear in the linear plot. (Despite this visual appearance, the fit quality is good.)

In \fig{fit_qq}, we show the fits for the $q\bar q$ and $qq'$ channels at NLO and NNLO. The $qq'$ channel is interesting since its LL power correction vanishes, and so it exhibits a different functional behavior. At NLO, the NLL power correction, $a_0$, reproduces exceptionally well the full nonsingular result. This provides strong motivation to also compute the $a_0$ coefficient for the other partonic channels, and to understand its universality, as its inclusion renders the power corrections for $N$-jettiness subtractions at NLO negligible. At NNLO, a similar pattern is observed as for the $gq$ channel albeit shifted by a logarithmic order, namely that the $a_2$ and $a_1$ coefficients have alternating signs and $a_1$ is larger than $a_2$. The $a_2$ coefficient alone therefore does not provide a particularly good approximation to the full nonsingular result, as is clearly seen in \fig{fit_qq} (bottom pannels).

In \figs{cumulantNLO}{cumulantNNLO} we plot the resulting integrated power corrections $\Delta\sigma(\tau_\cut)$ at NLO and NNLO. For the $gg$ channel, the expected scaling from \fig{scaling} is quite well reproduced, and removing the LL contribution yields a significant reduction of the power correction, up to an order of magnitude depending on the value of $\tau_\cut$. Interestingly, at NNLO it seems that the NLL contribution in the cumulant is quite small, so removing it only has a small effect. On the other hand, for the $gq$ channel, due to its color-suppressed LL contribution, the naive scaling does not apply and the full power correction is already determined by the NLL contribution. To obtain a further reduction, it is then necessary to also remove the NLL contributions. For the full nonsingular, the sum of the $gg+gq$ channels is dominated by $gg$ channel due to its larger LL contribution, which also means that removing the LL contribution yields the expected reduction in the power corrections at both NLO and NNLO. Interestingly, after removing the LL contribution, the remaining NLL (and beyond) power corrections are of similar size for the $gg$ and $gq$ channels. However, they are of opposite signs and partially cancel in the sum of both channels.

The $gg$ channel at NNLO, shown in \figs{fitNNLO}{cumulantNNLO}, clearly illustrates a potential pitfall if the power corrections in $N$-jettiness subtractions are not properly understood, namely the presence of false plateaus. At NNLO, the power corrections are cubic polynomials in $\ln{\tau_\cut}$, and hence will generically exhibit zero-crossings and extrema. For the $gg$ and $gq$ channels there is a maximum in the nonsingular spectrum around $\Tau_0=0.1-1$ GeV, which translates into a shallow maximum in the cumulant in the range $\tau_\cut \sim 10^{-3}-10^{-2}$. Since the maximum is very shallow, even with the high statistics that we have generated, it could easily appear as a plateau where changing the value of $\tau_\cut$ does not affect the cross section, leading to the false conclusion that the power corrections have become negligible. However, in this region the power corrections are still nonnegligible and amount to $1-2\%$ of the Born cross section. To avoid such false plateaus, even without generating data to significantly smaller values of $\Tau_0$, one can use the functional form of the power corrections, and extrapolate to $\Tau_0\to 0$. This was done in the recent calculation of $Z\gamma$ production at NNLO \cite{Campbell:2017aul} using $\Tau_0$.
We also note that such false plateaus can in principle appear in any global subtraction scheme.

The results of Ref.~\cite{Boughezal:2016zws} for the $gg\to H$ channel involve only derivatives of PDFs, and therefore do not agree with our results at NLO in \eq{NLOResult} and at NNLO in \eq{NNLOResult}, which involve both PDF derivatives and a constant term multiplying the PDFs. However, we do agree on the coefficients of the terms involving PDF derivatives. As discussed above, we have carefully checked our results by comparing to numerical results. The terms involving PDF derivatives contribute only $\approx 2/3$ of the full LL contribution at both NLO and NNLO, so our numerical cross checks confirm the presence of a constant term. At NLO, our fit result in table \ref{tab:LLresults} agrees within the $1\%$ fit uncertainty, clearly ruling out the result of Ref.~\cite{Boughezal:2016zws}. The same conclusion holds for the $q\bar q$ channel in the Drell-Yan case, see table I of Ref.~\cite{Moult:2016fqy}. On the other hand, without a detailed fit with multiple subleading coefficients this would be difficult to distinguish at NNLO, and indeed from \fig{cumulantNNLO} it is clear that a reduced value for the LL power correction would be preferred if the subleading power corrections were not carefully taken into account in the fit. Furthermore, Ref.~\cite{Boughezal:2016zws} used a hadronic definition (see \sec{discuss}), for which we have found that obtaining a reliable fit that is able to distinguish the contributions from different logarithmic powers is much harder due to the rapidity enhancement of additional subleading power corrections, which render the power expansion more poorly behaved.

%%%%%%%%%%%%%%%%%%%%%%%%%%%%%%%%%%%%%%%%%%%%%%%%%%%%%%%%%%%%%%%%%%%%%%%%%%%%%%%%
\section{Rapidity Dependence and Observable Definition}
\label{sec:discuss}
%%%%%%%%%%%%%%%%%%%%%%%%%%%%%%%%%%%%%%%%%%%%%%%%%%%%%%%%%%%%%%%%%%%%%%%%%%%%%%%%

\begin{figure*}[t]
\includegraphics[width=\columnwidth]{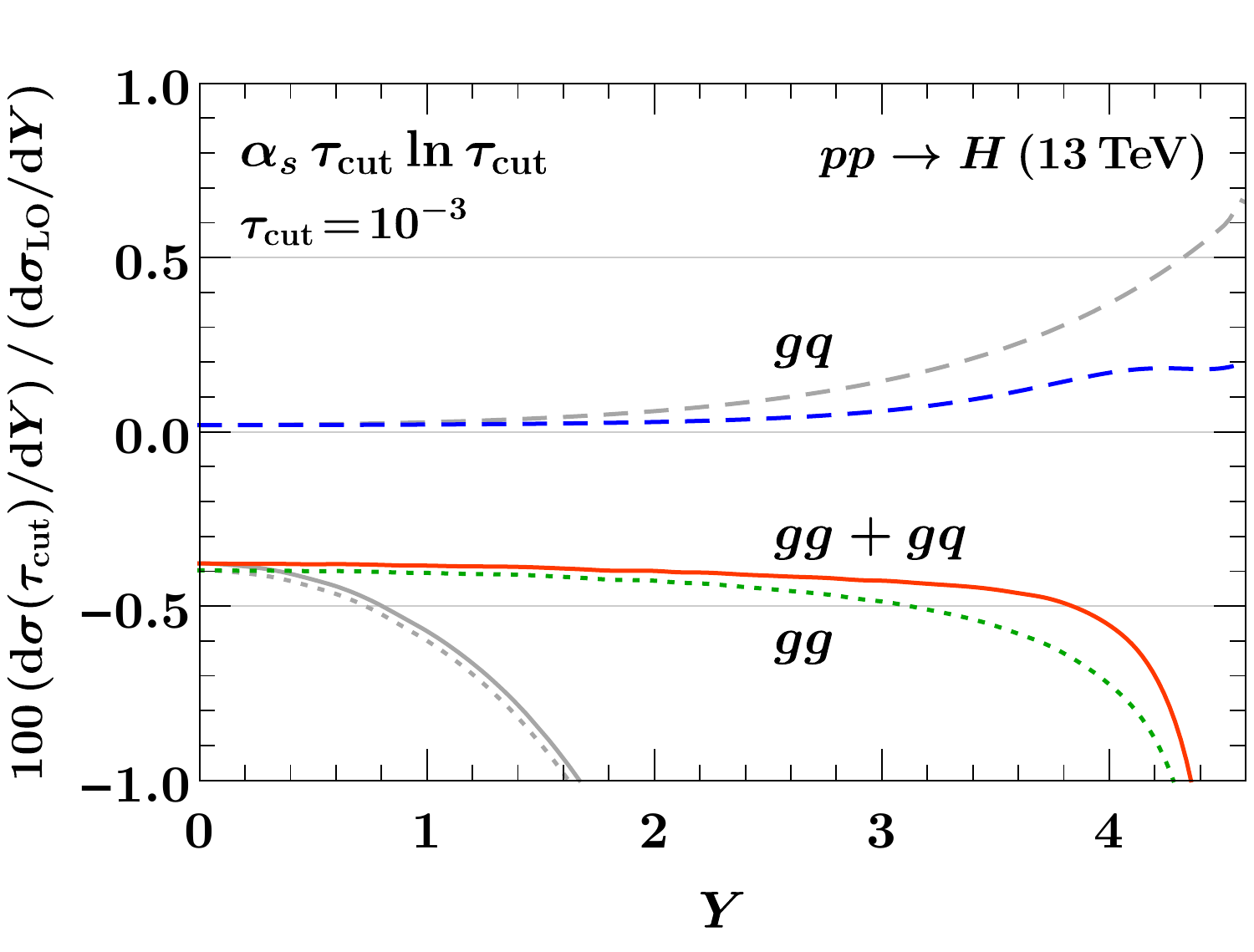}%
\hfill
\includegraphics[width=\columnwidth]{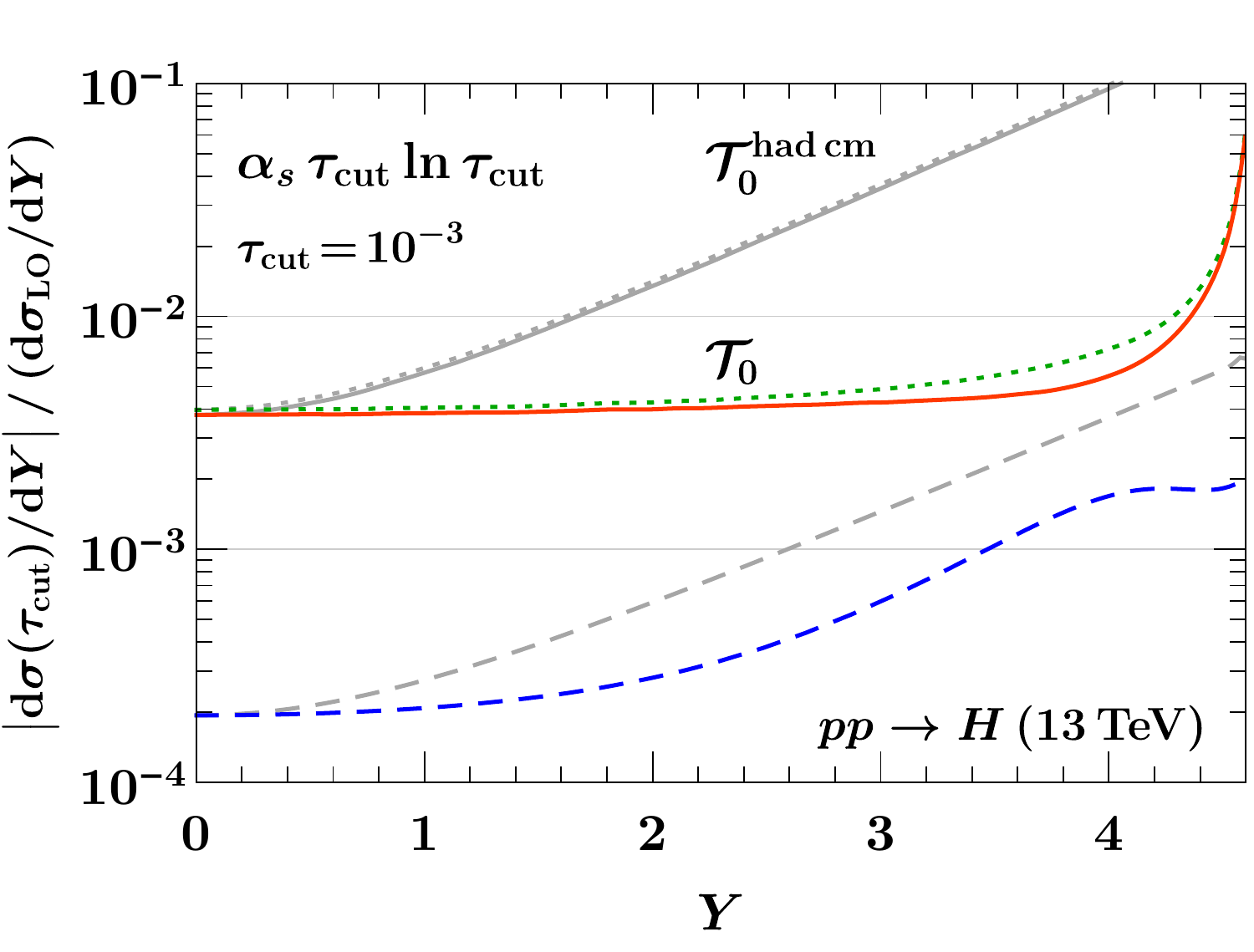}%
\\
\includegraphics[width=\columnwidth]{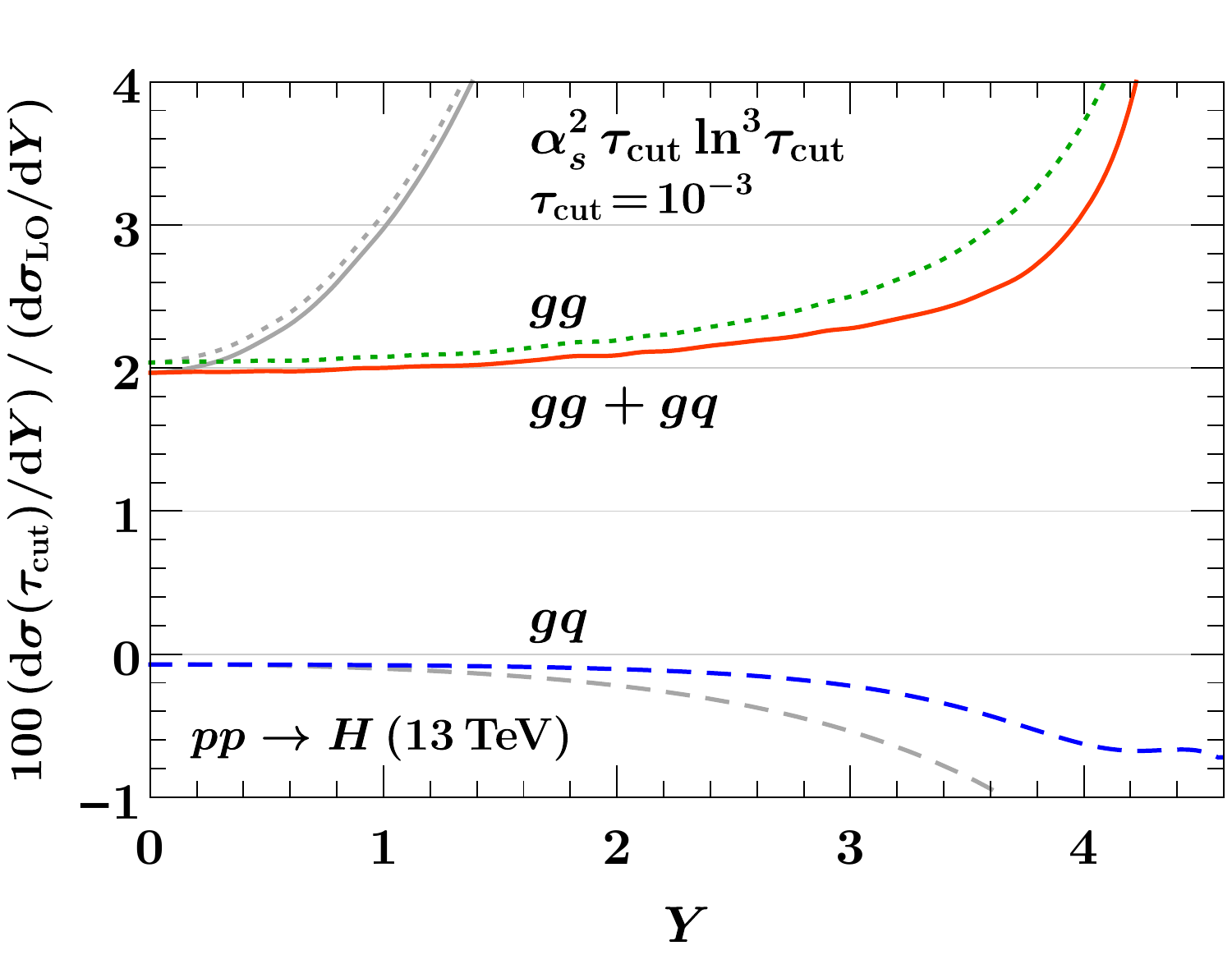}%
\hfill
\includegraphics[width=\columnwidth]{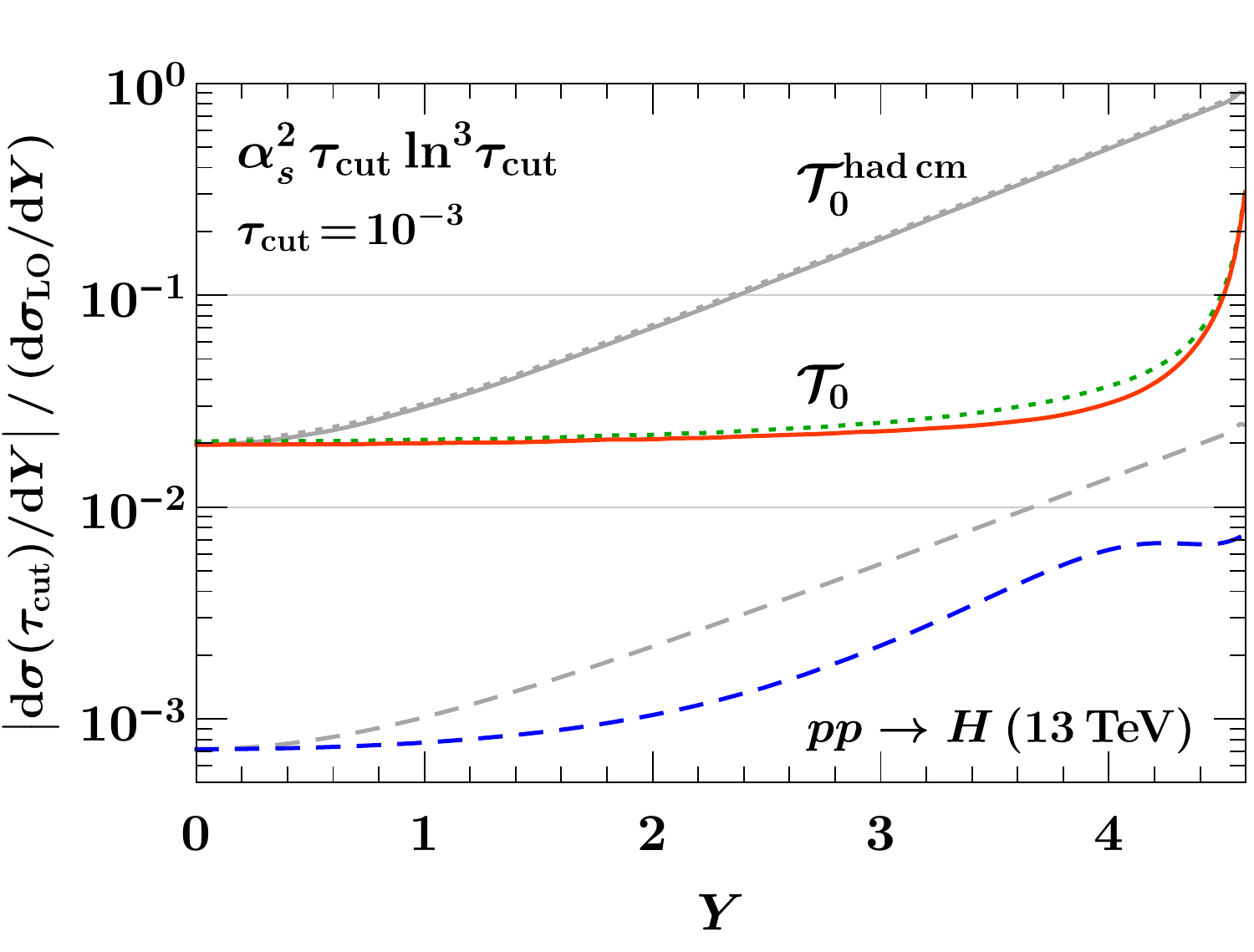}%
%%%
\caption{Leading-logarithmic power correction in the rapidity spectrum at $\ord{\alpha_s}$ (top row) and $\ord{\alpha_s^2}$ (bottom row), for the $gg$ (dotted), $gq$ (dashed), and $gg+gq$ (solid) channels using $\tau_\cut = 10^{-3}$. The colored curves show the standard definition of $\Tau_0$ in the leptonic frame, while the gray curves show the definition of $\Tau_0^\hadcm$ in the hadronic frame. The exponential growth of the power corrections for $\Tau_0^\hadcm$ are clearly visible.
The plots in the right column are identical to those in the left column, but show the absolute value on a logarithmic scale.}
\label{fig:rapidity}
\end{figure*}

One of the interesting observations in the analytic calculation of the power corrections for Drell-Yan production in Ref.~\cite{Moult:2016fqy} is that the structure of the power corrections depends sensitively on the definition of the $\Tau_0$ observable that is used. In particular, for the hadronic definition of $\Tau_0$, the power corrections grow exponentially with rapidity. This is also the case for $gg \to H$, where the LL power correction for the hadronic definition of $\Tau_0$ are given at NLO by
%%%
\begin{align} \label{eq:NLOResulthad}
\widetilde C_{gg, 1}^{(2,1)}(\xi_a, \xi_b)
&= 4C_A \Bigl[ e^Y \delta_a (\delta_b + \delta'_b) + e^{-Y} (\delta_a + \delta_a')\delta_b \Bigr]
 \,, \nn \\
\widetilde C_{qg, 1}^{(2,1)}(\xi_a, \xi_b)
&= -2 C_F \,e^{-Y}\delta_a \delta_b
 \,, \nn \\
\widetilde C_{gq, 1}^{(2,1)}(\xi_a, \xi_b)
&= -2 C_F \,e^{Y} \delta_a \delta_b
\,,\end{align}
%%%
and at NNLO by
%%%
\begin{align} \label{eq:NNLOResulthad}
\widetilde C_{gg, 3}^{(2,2)}(\xi_a, \xi_b)
&= -16 C_A^2 \Bigl[ e^Y \delta_a (\delta_b + \delta'_b) + e^{-Y} (\delta_a + \delta_a')\delta_b \Bigr]
, \nn \\
\widetilde C_{qg, 3}^{(2,2)}(\xi_a, \xi_b)
&=  4 C_F(C_F+C_A)e^{-Y} \delta_a \delta_b
\,, \nn \\
\widetilde C_{gq, 3}^{(2,2)}(\xi_a, \xi_b)
&=  4 C_F(C_F+C_A)e^{Y} \delta_a \delta_b
\,. \end{align}
%%%
Physically, this behavior can be understood as in Drell-Yan, either from the fact that for a boosted system the $0$-jettiness definition in the hadronic frame is no longer appropriately constraining soft and collinear radiation, or as  arising from the fact that the eikonal approximation is breaking down, so the soft emissions are sensitive to the momentum of the incoming partons.

To illustrate the importance of the leptonic definition of the $\Tau_0$ observable to yield power corrections that are approximately constant over phase space, in \fig{rapidity} we show the LL power correction as a function of the Higgs rapidity $Y$. The exponential growth of the power correction with $Y$ is clearly seen for the hadronic definition (gray curves labelled $\Tau_0^\hadcm$). For the leptonic definition (colored curves labelled $\Tau_0$), the power correction is flat in $Y$ as expected, except at very high rapidities, where the behavior of the PDFs enters. Since $N$-jettiness subtractions are meant for performing fully differential NNLO subtractions, the rapidity independence of power corrections is important to ensure that kinematic distributions are not distorted by missing power corrections. Although the leptonic definition we have used here relies on the presence of leptons, it can easily be generalized to the case of a fully hadronic final state, as described in Ref.~\cite{Moult:2016fqy}. We strongly recommend the use of such a definition in future applications of $N$-jettiness subtractions, particularly for more complicated processes where the power corrections have not yet been calculated. Indeed, the leptonic definition was used in the recent calculation of $Z\gamma$ production at NNLO \cite{Campbell:2017aul}.

%%%%%%%%%%%%%%%%%%%%%%%%%%%%%%%%%%%%%%%%%%%%%%%%%%%%%%%%%%%%%%%%%%%%%%%%%%%%%%%%
\section{Conclusions}
\label{sec:conclusions}
%%%%%%%%%%%%%%%%%%%%%%%%%%%%%%%%%%%%%%%%%%%%%%%%%%%%%%%%%%%%%%%%%%%%%%%%%%%%%%%%

In this paper, we have performed a detailed study of power corrections for $2$-jettiness in $H\to gg$, and $0$-jettiness (beam thrust) for $gg\to H$. We analytically computed the LL power correction at both NLO and NNLO, namely the $\alpha_s \ln \tau$ and $\alpha_s^2 \ln^3 \tau$ terms for both $2$-jettiness in $H\to gg$, and $0$-jettiness in $gg \to H$, for all partonic channels. We find partial agreement with the results of Ref.~\cite{Boughezal:2016zws}, as detailed in the body of the paper. The simplicity of the analytic results, and their close relation to those for quark-initiated processes suggests a degree of universality in the subleading soft and collinear limits, similar to that which is observed at leading power.

We confirmed our analytic results by comparing to the full nonsingular cross section obtained numerically from the full $H+1$ jet NLO calculation, and studied in detail the structure of the power corrections in the different partonic channels, $gg$, $gq$, and $q\bar q$. For the $gg$ channel, we found that the nonsingular corrections are well approximated by the LL term, and that the missing power corrections to the $0$-jettiness subtractions are reduced by up to an order of magnitude when including the analytically computed LL power corrections. On the other hand, for the $gq$ channel the LL power correction is small compared to the NLL contribution. However, the total cross section is dominated by the $gg$ channel at this level, so the inclusion of the LL power corrections overall significantly improves the performance of the subtractions. Knowing the LL contributions also allowed us to numerically extract the size of the NLL terms. Our results motivate the analytic calculation of the NLL terms to fully understand their nontrivial structure and provide further improved control of the power corrections in all partonic channels. We also computed the NLL power correction for the $q\bar q$ channel, which does not have a LL power correction, giving a first hint at their structure. 

The numerical results for the power corrections also allowed us to study their rapidity dependence. As for the case of Drell-Yan, the power corrections exhibit an exponential growth with rapidity using the hadronic definition for $\Tau_0$, while they are nearly flat as a function of rapidity using the standard definition that takes into account the boost of the Born system.

Due to the importance of NNLO calculations for Run 2 of the LHC, it is essential to further improve the numerical efficiency of $N$-jettiness subtractions through a better understanding of the power corrections. This includes the calculation of the subleading logarithms for color-singlet production as well as the calculation of the leading-logarithmic power corrections for processes involving jets in the final state. We plan to address these directions in future work.

%%%%%%%%%%%%%%%%%%%%%%%%%%%%%%%%%%%%%%%%%%%%%%%%%%%%%%%%%%%%%%%%%%%%%%%%%%%%%%%%
\begin{acknowledgments}

We thank Radja Boughezal, Xiaohui Liu and Frank Petriello for correspondence regarding the results of \cite{Boughezal:2016zws}.
We thank the Erwin Schr\"odinger Institute and the organizers of the ``Challenges and Concepts for Field Theory and Applications in the Era of LHC Run-2'' workshop, as well as the Aspen Center for Theoretical Physics, which is supported by National Science Foundation grant PHY-1607611, for hospitality and support while portions of this work were completed.
This work was supported in part by the Office of Nuclear Physics of the U.S.
Department of Energy under Contract No. DE-SC0011090, by the Office of High Energy Physics of the U.S. Department of Energy under Contract No. DE-AC02-05CH11231,
by the DFG Emmy-Noether Grant No. TA 867/1-1,
the Simons Foundation Investigator Grant No. 327942, and the LDRD Program of LBNL.

\end{acknowledgments}
%%%%%%%%%%%%%%%%%%%%%%%%%%%%%%%%%%%%%%%%%%%%%%%%%%%%%%%%%%%%%%%%%%%%%%%%%%%%%%%%

\bibliography{../subleading}

\end{document}